\documentclass[twocolumn,amsmath,amssymb,superscriptaddress,prl,floatfix,footinbib]{revtex4-1}
\usepackage[]{graphicx}
\usepackage[usenames,dvipsnames]{color}
\usepackage{soul}
\usepackage{bm}

\usepackage{times}
\usepackage{amsfonts}
\usepackage{mathrsfs}
\usepackage{graphicx}
\usepackage{dcolumn}
\usepackage{bm}
\usepackage{color}
\usepackage{xcolor}

\usepackage[colorlinks,bookmarks=false,citecolor=blue,linkcolor=red,urlcolor=blue]{hyperref}
\usepackage{multirow}
\usepackage{physics}

\begin{document}

\title{Deep Learning the Functional Renormalization Group}

\author{Domenico Di Sante}
\affiliation{Department of Physics and Astronomy, University of Bologna, 40127 Bologna, Italy}\email{domenico.disante@unibo.it}
\affiliation{Center for Computational Quantum Physics, Flatiron Institute, 162 5th Avenue, New York, NY 10010, USA}

\author{Matija Medvidović}
\affiliation{Center for Computational Quantum Physics, Flatiron Institute, 162 5th Avenue, New York, NY 10010, USA}
\affiliation{Department of Physics, Columbia University, New York, NY 10027, USA}

\author{Alessandro Toschi}
\affiliation{Institute of Solid State Physics, TU Wien, A-1040 Vienna, Austria}

\author{Giorgio Sangiovanni}
\affiliation{Institut f\"ur Theoretische Physik und Astrophysik and W\"urzburg-Dresden Cluster of Excellence ct.qmat, Universit\"at W\"urzburg, 97074 W\"urzburg, Germany}

\author{Cesare Franchini}
\affiliation{Department of Physics and Astronomy, University of Bologna, 40127 Bologna, Italy}
\affiliation{University of Vienna, Faculty of Physics and Center for Computational Materials Science, A-1090 Vienna, Austria}

\author{Anirvan M. Sengupta}
\affiliation{Center for Computational Quantum Physics, Flatiron Institute, 162 5th Avenue, New York, NY 10010, USA}
\affiliation{Center for Computational Mathematics, Flatiron Institute, 162 5th Avenue, New York, NY 10010, USA}
\affiliation{Department of Physics and Astronomy, Rutgers University, Piscataway, NJ 08854, USA}

\author{Andrew J. Millis}
\affiliation{Center for Computational Quantum Physics, Flatiron Institute, 162 5th Avenue, New York, NY 10010, USA}
\affiliation{Department of Physics, Columbia University, New York, NY 10027, USA}

\date{\today}

\begin{abstract}
    We perform a data-driven dimensionality reduction of the scale-dependent 4-point vertex function characterizing the functional Renormalization Group (fRG) flow for the widely studied two-dimensional $t - t'$ Hubbard model on the square lattice. We demonstrate that a deep learning architecture based on a Neural Ordinary Differential Equation solver in a low-dimensional latent space efficiently learns the fRG dynamics that delineates the various magnetic and $d$-wave superconducting regimes of the Hubbard model. We further present a Dynamic Mode Decomposition analysis that confirms that a small number of modes are indeed sufficient to capture the fRG dynamics. Our work demonstrates the possibility of using artificial intelligence to extract compact representations of the 4-point vertex functions for correlated electrons, a goal of utmost importance for the success of cutting-edge quantum field theoretical methods for tackling the many-electron problem.
\end{abstract}

\maketitle

{\it Introduction --}
Interacting electron systems exhibit a rich variety of distinct phenomena at different energy and temperature scales. Upon lowering these scales, new effective degrees of freedom and collective behaviors emerge, typically including competing spin, charge and pairing fluctuations. The difficulties inherent in treating these competing, scale-dependent phenomena on an equal footing represent one of the major obstacles to the numerical solution of theoretical models.

The renormalization group (RG) provides a powerful approach to these problems~\cite{wilson1974renormalization, WilsonRMP, Fisher1983, ShankarRMP, polchinski1992effective}. The characteristic of RG of keeping only the relevant information, as a scale parameter is reduced, makes it a valuable tool to deal with interacting fermions. In its {\it exact} or {\it functional} (``fRG") form, the RG is formulated as an exact functional flow equation which, as a function of a continuously decreasing energy scale, provides an effective action description of a microscopic model~\cite{Wetterich,Morris,SalmhoferPTP2001,KopietzPRB2001}.

In quantum condensed matter physics the fRG has been used to study model systems such as the two-dimensional (2D) Hubbard model and its extensions~\cite{ZanchiPRB2000, HalbothPRB2000, HalbothPRL2000, HonerkampPRB2001, HonerkampSalmhoferPRL, HonerkampSalmhoferPRB,Hille2020,Hille2020B,Vilardi2020}.
Applications of fRG to real materials have thus far remained sporadic, requiring considerable numerical effort to incorporate realistic band-dispersions, multi-orbital characters and realistic interactions. Successful applications, including the study of SC and competing phases in iron-based compounds~\cite{WangPRL2009, ThomalePRB2009, ThomalePRL2011a, ThomalePRL2011b}, cobaltates~\cite{KieselPRL2013}, doped and twisted-bilayer graphene~\cite{HonerkampPRL2008, KieselPRB2012, KennesPRB2018}, buckled Dirac semimetals, doped topological insulators~\cite{GeMoS2fRG, WuPRB2019} and quantum spin liquid phases in frustrated antiferromagnets~\cite{Chillal2020,BuessenPRL2018}, show the potential of this approach, but the underlying computational complexity suggests a simplification of the fRG approach would be desirable.

In the standard RG procedure, say, for a $(4-\epsilon)$-dimensional $\phi^4$ field theory with $\epsilon$-expansion~\cite{wilson1974renormalization}, the effective action of the theory takes a simple form parametrized by a small number of coupling constants and the functional flow equations collapse to a small set of coupled differential equations describing the flow of these coupling constants. In contrast, the common formulation of fermionic fRG keeps track of the entire frequency/momentum-dependence of the interaction vertexes during the flow~\cite{HonerkampPRB2001,KopietzPRB2001}. Thus, the apparent dimensionality space of `couplings' is high, although schemes for finding compressed representations have been investigated ~\cite{NilsPRB2020,Parcollet,Hille2020,Hille2020B,HusemannPRB2009,WangPRB2012,LichtensteinTUfRG}. In this paper we present results from a data-driven approach indicating that this apparent high dimensionality of the vertex function can be, in some cases, illusory.

In the context of high-dimensional data, the advent of machine learning (ML) techniques and data-driven approaches applied to many body quantum physics has triggered enormous interest~\cite{CarleoRMP}. ML ideas have been applied so far to several categories of methods for interacting electron systems, including density functional theory (DFT)~\cite{BurkePRL2012, BurkeNatComm2017, GeorgesPRL, KirkpatrickScience2021}, the Anderson impurity model~\cite{MillisPRB2014}, quantum-embedding and dynamical mean field theory (DMFT)~\cite{LanataPRR2021,LeeDMFT2019, sheridan2021datadriven}, and the numerical renormalization group (NRG)~\cite{LeiWang,RigoNRG}. Interacting spins models have also been studied~\cite{CarleoScience}. In fact, the ability of neural networks to approximate a very large class of functions promoted the use of deep net architecture as a new numerical tool for solving the quantum many-body problem.


The essential object in the fRG is the vertex function $\text{V}(\text{k}_1,\text{k}_2,\text{k}_3)$,  whose description, in principle, requires the computation and storage of a function of three continuous momentum variables. By studying a particular theoretical model, the two dimensional  $t-t^\prime-U$ Hubbard model, believed relevant for cuprates and wide classes of organic conductors, we show that a lower dimensional representation can capture the fRG flow of the apparently high-dimensional vertex functions. To learn this simpler representation, we use a neural network architecture  known as Neural Ordinary Differential Equations (NODE) \cite{Chen2018}.
The latent variable space of the NODE provides us with insight into the low-dimensional structure of the fRG flow. In order to further investigate the reason for this simplicity, we apply Dynamic Mode Decomposition (DMD)~\cite{schmid_dynamic_2010, rowley2009spectral}, a complementary dimensionality reduction scheme that is specifically tailored for dynamical systems. We observe that a small number of modes are able to approximately capture the fRG dynamics of this model. In other words, the key to success of the method is that the 
momentum-dependence of the vertex can be approximated by combining a small number of patterns characteristic of the different competing many-body phases.

Thus our ML approach to fRG achieves what reduced order models~\cite{Quarteroni} wish to accomplish. However, we do not apply predetermined ans\"atze or make simplifying approximations, potentially discarding some relevant information in the vertex function. Instead, we let data guide the choice of the lower-dimensional representation.

{\it The fRG ground states of the Hubbard model --} The microscopic Hamiltonian we consider is
\begin{equation}
\label{Eq1}
    H = -t \sum_\text{nn,s} c^{\dag}_{i,s}c_{j,s} - t' \sum_\text{nnn,s} c^{\dag}_{i,s}c_{j,s} + U \sum_i n_{i,\uparrow}n_{i,\downarrow}
\end{equation}

\noindent with hopping amplitudes $t$ and $t'$ between nearest neighbours (nn) and next-nearest neighbours (nnn) on the 2D square lattice, and onsite Coulomb repulsion $U$. The 2-particle properties of this model are investigated through the temperature-flow one-loop fRG scheme~\cite{HonerkampSalmhoferPRL,HonerkampSalmhoferPRB}, where the RG flow of $\text{V}^\Lambda(\text{k}_1,\text{k}_2,\text{k}_3)$ is
\begin{equation}
\label{Eq2}
    \frac{\dd \text{V}^\Lambda}{\dd \Lambda} = \text{V}^\Lambda \, \circ \text{L}^\Lambda \, \circ \text{V}^\Lambda \,
\end{equation}

\begin{figure}[!t]
\centering
\includegraphics[width=\columnwidth,angle=0,clip=true]{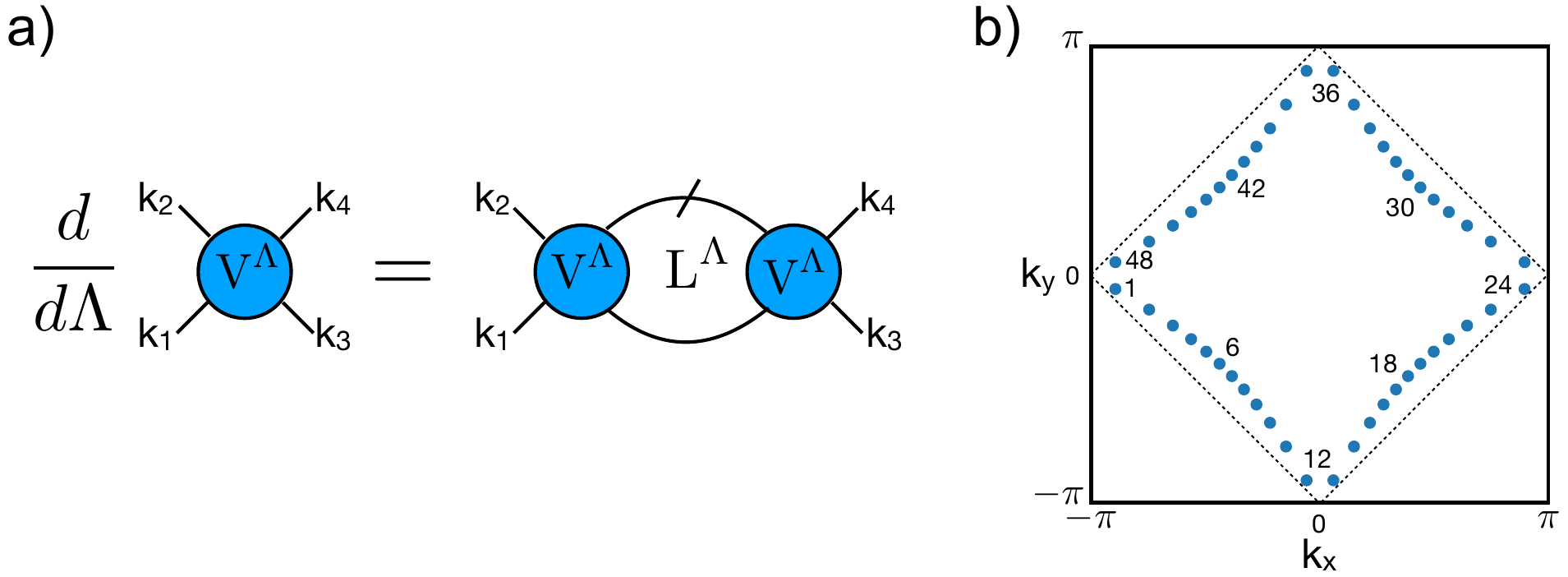}
\caption{
    a) Diagrammatic representation of the one-loop fRG flow equation for the 2-particle vertex function $\text{V}^{\Lambda}$.
    b) Fermi surface (FS) of the $t-t'$ tight-binding model for $t'=-0.25t$. The blue points indicate  the 48 momenta used to patch the FS. The black dashed lines are the Umklapp surface of perfect nesting at $t^\prime=0$.
}
\label{fig1}
\end{figure}

\noindent (see Fig.~\ref{fig1}a), with the RG scale $\Lambda$ given by the temperature $T$, and $\circ$ defining integration over the internal degrees of freedom. Comprehensive reviews of the fRG scheme applied to fermionic problems are found in Refs.~\cite{MetznerRMP, PlattReview}; Ref.~\cite{KopietzBook} gives an excellent pedagogical introduction. In essence, for spin-rotation invariant systems, the right-hand-side of Eq.~\ref{Eq2} splits into the sum of three contributions, which describe the particle-particle, direct particle-hole and crossed particle-hole channels~\cite{HonerkampPRB2001} necessary to account on similar footing for the SC and density-wave instabilities. $\text{L}^\Lambda$ is a scale-dependent loop kernel that contains information on the single-particle properties of the microscopic model~\cite{note1}.

Neglecting the frequency dependences of the vertex couplings, which have a negative scaling dimension (irrelevant couplings) under the RG flow~\cite{note4}, but keeping a full momentum description in terms of a discrete set of $N_k$ wave vectors on the Fermi surface (FS), Eq.~\ref{Eq2} is recasted into a set of $N_k^3$ coupled ordinary differential equations (ODE). The solution to this problem, with initial conditions $\text{V}^{\Lambda_0}(\text{k}_1,\text{k}_2,\text{k}_3) = U$ when $\Lambda_0 = 8t$ is the bandwidth, yields the gradual evolution of the 2-particle vertex function $\text{V}^\Lambda(\text{k}_1,\text{k}_2,\text{k}_3)$ as $\Lambda \rightarrow 0$ approaches the FS. For a typical $N_k = 48$ FS discretization, as depicted in Fig.~\ref{fig1}b), the complexity of Eq.~\ref{Eq2} already amounts to more than $10^5$ coupled ODE.

\begin{figure}[!t]
\centering
\includegraphics[width=\columnwidth,angle=0,clip=true]{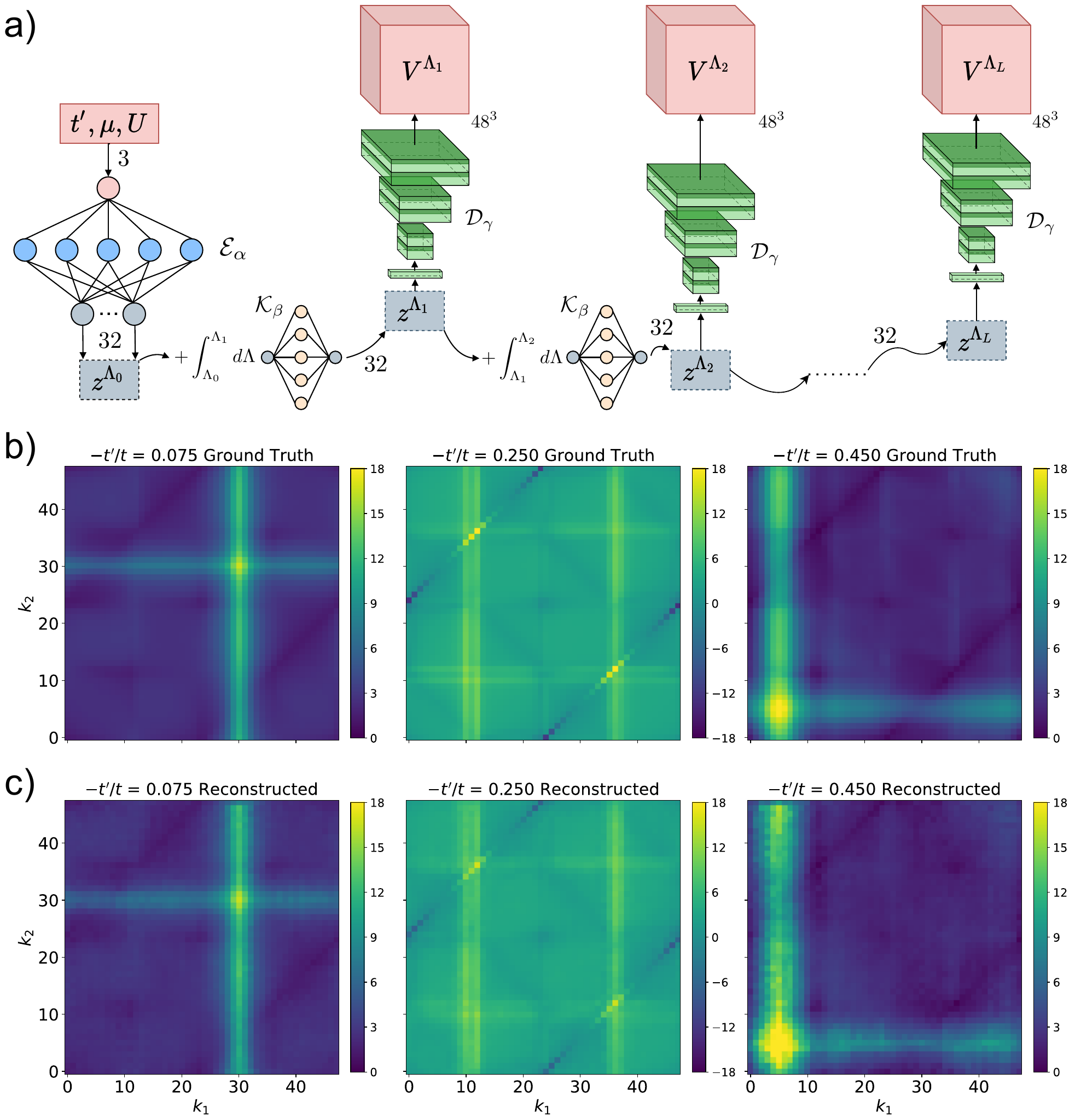}
\caption{
    a) The deep learning architecture defined by Eqs.~\ref{eq_NODE} and detailed in Ref.~\cite{SuppMatt}.
    b) False-color representation of amplitude of vertex function $\text{V}^{\Lambda}(\text{k}_1,\text{k}_2,\text{k}_3)$ at a late stage of the renormalization process, for different representative initial conditions. To represent a function of three momenta we fix the first outgoing wave vector $\text{k}_3$ and present the vertex as a function of the two independent incoming vectors.
    Following Ref.~\cite{HonerkampSalmhoferPRB}, (left) V for $t^\prime=-0.075t$ with $\text{k}_3$ fixed at patch 6; (central) V for $t'=-0.25t$ and $\text{k}_3$ is fixed at patch 25 to emphasize the diagonal features important for the Cooper channel; (right) V for  $t'=-0.45t$ and $\text{k}_3$ fixed at patch 6.
    These choices correspond to AFM, $d$-wave SC and FM instabilities respectively~\cite{HonerkampSalmhoferPRL}
    c) Same as b) but for the predicted data $\hat{\text{V}}^{\Lambda}(\text{k}_1,\text{k}_2,\text{k}_3)$. We highlight that these data belong to the test set.
}
\label{figNODE_1}
\end{figure}

When Eq.~\ref{Eq2} is numerically solved at varying $t'$ for chemical potential fixed at the van Hove filling ($\mu = 4t'$) and at weak-coupling $U=3t$, the Hubbard model in Eq.~\ref{Eq1} experiences three different regimes~\cite{HonerkampSalmhoferPRL}: i) The first regime is close to half-filling, with $t' > -0.2t$ and dominant AF scattering processes between FS regions connected by wave vectors $\sim (\pi,\pi)$. These can be seen in Fig.~\ref{figNODE_1}b) (left) as bright features corresponding to repulsive couplings on the line $\text{k}_2-\text{k}_3 \sim (\pi,\pi)$. ii) Further decreasing $t'$, $d$-wave SC takes over, with the dominant $d_{x^2-y^2}$ symmetry of the pairing scattering, as seen from the sign profile of the diagonal features of Fig.~\ref{figNODE_1}b) (central). iii) After a quantum critical point for $t' \sim -0.33t$, scattering processes with small momentum transfer $\text{k}_2-\text{k}_3 \sim (0,0)$ dominate, see bright features in Fig.~\ref{figNODE_1}b) (right), leading to a change of ground state from $d$-wave singlet SC to ferromagnetism (FM).

{\it The Deep Learning fRG: results and interpretation --}
By inspecting the $\mathcal{O}(10^5)$ couplings of the 2-particle vertex functions of Fig.~\ref{figNODE_1}b) just before the fRG flow runs to strong coupling and the one-loop approximation breaks down, we recognize that many of them either have remained nearly constant or have become vanishingly small under the RG flow. Only {\it few} of them have grown positively or negatively (bright features) under the RG evolution~\cite{note4}. However, as mentioned before,  contrary to the standard RG procedure for traditional critical phenomena~\cite{Fisher1983}, fermionic fRG does not discard any coupling in the vertex $\text{V}^{\Lambda}(\text{k}_1,\text{k}_2,\text{k}_3)$ during the flow. Our approach is to find a simpler representation in a data-driven manner, using the power of neural nets to find useful features.

In recent years, there have been many developments in utilizing neural networks for predicting sequence data. These range from conventional Recurrent Neural Network (RNN), gated RNNs, like the Long Short-Term Memory (LSTM) and those using the gated recurrent unit (GRU)~\cite{goodfellow2016deep}, all the way to Encoder-Decoder with attention~\cite{vaswani2017attention}. Since we are interested in finding latent variables whose dynamics itself is governed by an ODE, the natural candidate is a flexible dimensionality reduction scheme based on the Parameterized NODE architecture~\cite{Chen2018,Lee2021}. The method, sketched in Fig.~\ref{figNODE_1}a), focuses on three deep neural networks -- the encoder $\mathcal{E}$, the NODE $\mathcal{K}$ and the decoder $\mathcal{D}$. The complete action of our model is defined by:
\begin{equation}
\label{eq_NODE}
    \vb{z}^{\Lambda _0} = \mathcal{E} _\alpha (t', U, \mu) \ ;\ \frac{\dd \vb{z}^\Lambda}{\dd \Lambda} = \mathcal{K} _\beta (\vb{z}^\Lambda) \ ;\ \hat{\text{V}}^\Lambda = \mathcal{D} _\gamma (\vb{z}^\Lambda)
\end{equation}

   
\noindent where $\alpha$, $\beta$, $\gamma$ are parameter sets corresponding to each neural network (details are found in Ref.~\cite{SuppMatt}, and our PyTorch implementation, NeuralFRG, is at Ref.~\cite{NeuralFRG}). The ground truth data are generated by solving the fRG problem in Eq.~\ref{Eq2} for 35 values of $t'$ in the range $0 \leq -t'/t < 0.5$ (and $U=3t$), and storing for each $t'$ the whole vertex dynamics, for a total of $\sim7$ thousands collected vertices, each with $\mathcal{O}(N_k^3 \sim 10^5)$ elements.

\begin{figure}[!b]
\centering
\includegraphics[width=\columnwidth,angle=0,clip=true]{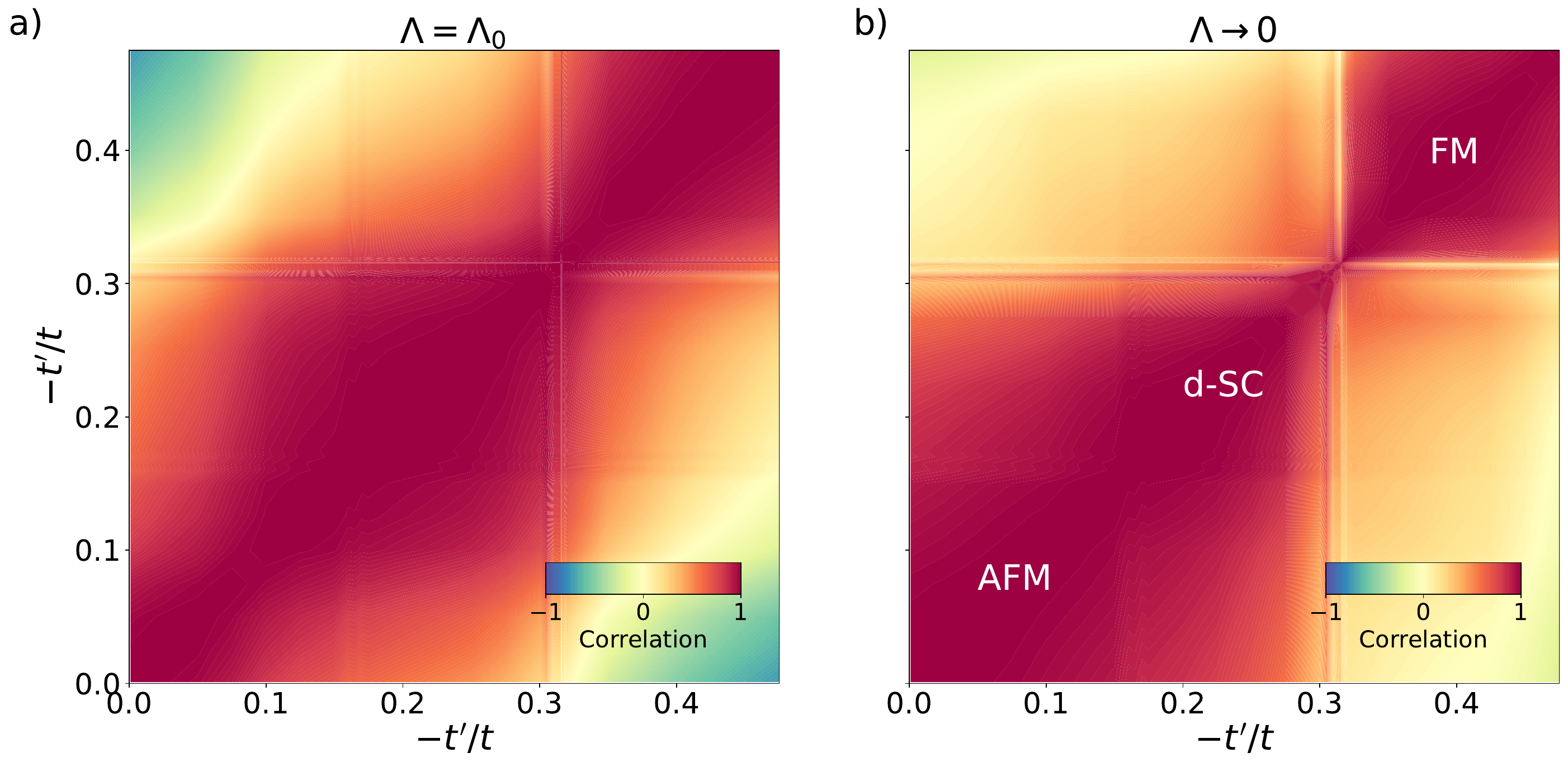}
\caption{
    Correlation matrix of the latent vectors $\vb{z}^\Lambda$ at different $t'$ values for $\Lambda = \Lambda_0$ (a) and $\Lambda \rightarrow 0$ (b), respectively. The correlation matrix is defined as the scalar product kernel $[\vb{\bar{z}}(t'_1) \cdot \vb{\bar{z}}(t'_2)]$ between normalized latent variables $\vb{\bar{z}}$, where $t'_1$ and $t'_2$ are any of the $0 \leq -t'/t < 0.5$. Red (blue) features correspond to a high (low) degree of statistical correlation.
}
\label{figNODE_2_part1}
\end{figure}

The encoder maps the Hubbard model parameters and fRG initial condition to a low-dimensional latent representation $\vb{z}^{\Lambda_0}$ of drastically smaller dimension than $N_k^3$. All the results here are obtained with a latent space dimension of 32, but are robust against the use of either smaller or larger values, as we show in Ref.~\cite{SuppMatt}. The NODE then defines a differential equation propagation rule for latent variables in $\Lambda$. Finally, at each step of the flow, a decoder network is employed to map the latent representation $\vb{z}^{\Lambda}$ to a reconstructed 4-point vertex function $\hat{\text{V}}^{\Lambda}(\text{k}_1,\text{k}_2,\text{k}_3)$. We use a modified version of mean squared error (MSE, Ref.~\cite{SuppMatt}) between $\hat{\text{V}}^\Lambda$ and $\text{V}^\Lambda$, in conjunction with a gradient-based optimizer \cite{Kingma2015}. All three networks, $\mathcal{E} _\alpha,\mathcal{K} _\beta$, and, $\mathcal{D} _\gamma$, are optimized simultaneously. 

We find that the learned dynamics successfully captures the final instability for the entire range of next-nearest neighbour couplings $t'$. Satisfactory prediction of the qualitative features of the vertex data is achieved in the limit $\Lambda \rightarrow 0$, as presented in Fig.~\ref{figNODE_1}c). More interestingly, Fig.~\ref{figNODE_2_part1} shows that, during the fRG dynamics in the latent space, three highly statistically correlated latent space representations $\vb{z}$ emerge as a learned feature of the NODE neural network. At $\Lambda = \Lambda_0$, a first classification task is performed by the encoder $\mathcal{E} _\alpha$, which produces highly-correlated latent variables according to the value of $t'$ (Fig.~\ref{figNODE_2_part1}a). The NODE $\mathcal{K} _\beta$ takes it over to finite RG-time $\ln\Lambda_0/\Lambda$, and during the final stages of the fRG evolution in $\Lambda$, three markedly correlated areas appear, as shown in Fig.~\ref{figNODE_2_part1}b).  

\begin{figure}[!t]
\centering
\includegraphics[width=\columnwidth,angle=0,clip=true]{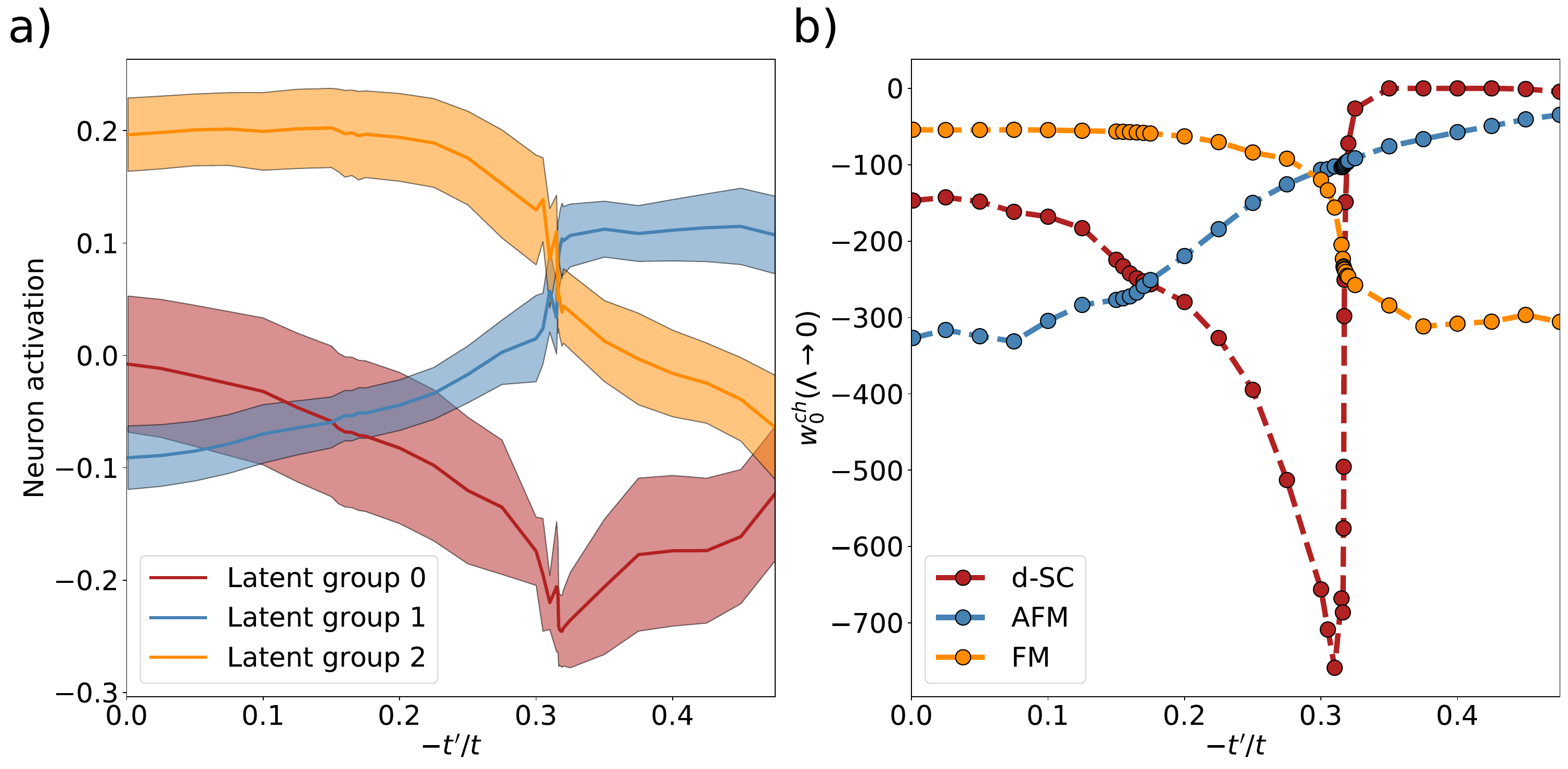}
\caption{
    a) Three-groups K-Means clustering~\cite{Kmeans} of neuron activation in $\mathcal{K} _\beta$ as a function of $t'$ for $\Lambda \rightarrow 0$.
    b) Evolution of the three channel leading eigenvalues $w_0^{ch}(\Lambda)$ as a function of $t'$ before the breaking of the one-loop approximation.
}
\label{figNODE_2_part2}
\end{figure}

The boundaries of these three regions roughly coincide with the values of $t'$ at which the fRG predicts a change in the leading two-particle instability~\cite{HonerkampSalmhoferPRL}. It is also interesting to notice that while the AFM and $d-$wave SC areas show similar normalized $\vb{\bar{z}}$ and are thus well-aligned in the latent space (the scalar product kernel $[\vb{\bar{z}}(t'_1) \cdot \vb{\bar{z}}(t'_2)] \sim 1$), reflecting their common origin in the dominant spin-fluctuations, the FM region stands on its own, separated from the other two phases by either a quantum critical point (in the fRG framework~\cite{HonerkampSalmhoferPRL}) or a first order transition (in the Hartree-Fock treatment~\cite{LinPRB1987}).  The neural network distinguishes between these three many-body regimes by learning specific low-dimensional hidden representations. This is accomplished by activating three different groups of neurons in $\mathcal{K} _\beta$ as a function of $t'$, as shown in Fig.~\ref{figNODE_2_part2}a). Each instability ground state corresponds indeed to a specific pattern of active neurons. This is manifestly evident when the neuron activation pattern of Fig.~\ref{figNODE_2_part2}a) is contrasted to Fig.~\ref{figNODE_2_part2}b), where we show the dependence on $t'$ of the most negative eigenvalues $w_0^{ch}(\Lambda)$ of the fRG channel-couplings $W^{\Lambda,ch}(\text{k}_1,\text{k}_2) = \sum_i w_i^{ch}(\Lambda)f_i^{ch}(\text{k}_1)^*f_i^{ch}(\text{k}_2)$~\cite{PlattReview,SuppMatt}, with channels $ch = \text{AFM, SC, FM}$ and $f_i^{ch}(\text{k})$ lattice harmonics transforming as an irreducible representation of the symmetry group of $W^{\Lambda,ch}(\text{k}_1,\text{k}_2)$. These leading eigenvalues are the ones associated with the highest ordering temperature $T_c$ for their specific channel~\cite{PlattReview}.

{\it The Dynamic Mode Decomposition (DMD) --}
The success of the NODE-based neural network in finding a relatively low-dimensional hidden representation prompts us to uncover possible simplicity hidden in the collection of the  dynamic trajectories of the vertex functions themselves. To explore this possibility, we apply conventional DMD~\cite{schmid_dynamic_2010}, without any specialized kernel~\cite{williams2014kernel}. Given a time series of data, DMD computes a set of modes each of which is associated with an eigenvalue on the complex plane, approximating eigenvalues and eigenvectors of the Koopman operator~\cite{rowley2009spectral}. Although the mathematical procedure for identifying the DMD modes and eigenvalues is purely linear, the dynamic itself can be nonlinear.

\begin{figure}[!t]
\centering
\includegraphics[width=\columnwidth,angle=0,clip=true]{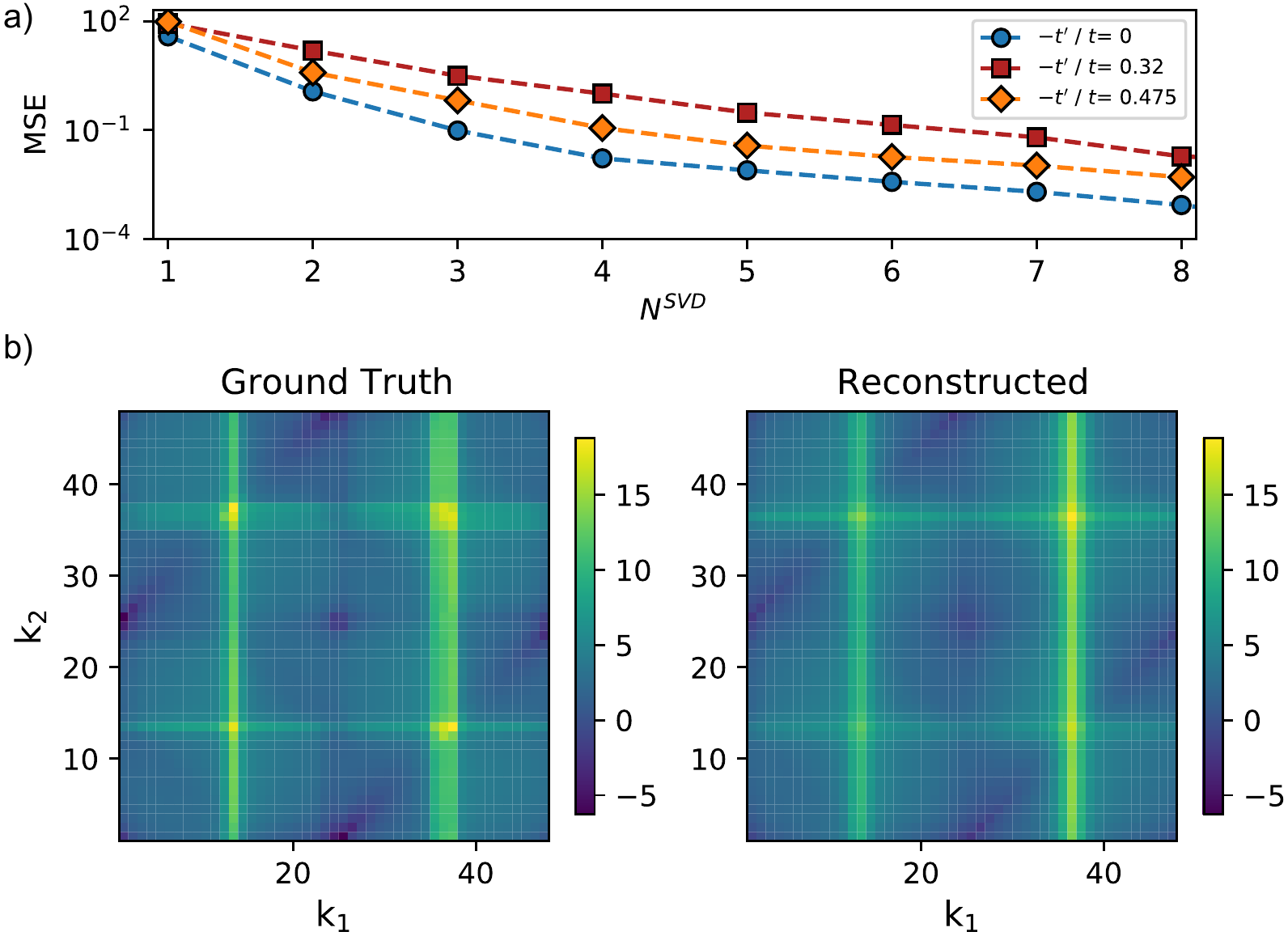}
\caption{
    a) MSEs from the DMD reconstructed fRG dynamics for $-t'/t = 0$ (AFM), $-t'/t = 0.32$ $d$-wave SC and $-t'/t = 0.475$ FM regimes as a function of the number of singular values $N^{SVD}$ included in the SVD scheme.
    b) Comparison between unseen ground truth $\text{V}^{\Lambda}(\text{k}_1,\text{k}_2,\text{k}_3)$ and DMD reconstructed data $\hat{\text{V}}^{\Lambda}(\text{k}_1,\text{k}_2,\text{k}_3)$ for $-t'/t = 0.17$ and first outgoing wave vector $\text{k}_3$ fixed at patch 1.
}
\label{fig3}
\end{figure}

For each individual fRG flow for $0 \leq -t'/t < 0.5$, we collected the 4-point vertex functions in the form of a snapshot sequence $\mathcal{V}_0^N = \{ \text{V}^{\Lambda_0},\text{V}^{\Lambda_1},...,\text{V}^{\Lambda_N}  \}$. These snapshots are assumed to be related via a linear mapping that defines a linear dynamical system in the Hilbert space of functions that best approximates the non-linear dynamics of the data~\cite{KutzBook} 
\begin{equation}
\label{Eq5}
    \text{V}^{\Lambda_{i+1}} = \mathcal{A} \text{V}^{\Lambda_{i}} \, ,
\end{equation}

\noindent with the Koopman operator $\mathcal{A}$ having the prohibitive dimension $N_k^3 \times N_k^3$. Eigenvalues and eigenvectors of $\mathcal{A}$ are referred to as the DMD eigenvalues and DMD modes respectively. Since a direct eigen-decomposition of $\mathcal{A}$ is unfeasible, the problem can be made tractable with the help of a Singular Value Decomposition (SVD) of $\mathcal{V}_0^{N-1}$~\cite{schmid_dynamic_2010,SuppMatt}.

One can select a restricted number $N^\text{SVD}$ of largest singular values of $\mathcal{V}_0^{N-1}$ performing a low-rank truncation, and optimally reconstruct the flow dynamics $\hat{\text{V}}^{\Lambda}(\text{k}_1,\text{k}_2,\text{k}_3)$ only using the leading $N^\text{SVD}$ DMD modes (see Ref.~\cite{Jovanovic2014} for details about the optimal reconstruction procedure used here). In Fig.~\ref{fig3}a), for exemplary cases of the three instability regimes, we show the MSE between the reconstructed data $\hat{\text{V}}^{\Lambda}$ and the ground truth data $\text{V}^{\Lambda}$, which rapidly decays with the number of included DMD modes.

Interestingly, one can also use several different fRG flows in the same snapshot sequence and perform the DMD analysis. This way, the DMD modes inherit features from all included 4-point vertex functions. We choose $-t'/t = 0$ (AFM), $-0.315$ ($d-$wave SC), $-0.475$ (FM) in order to span all three instability regimes, and reconstruct the fRG flow even on unseen data. Fig.~\ref{fig3}b) shows, for instance, the reconstructed data for $-t'/t = 0.17$ and $N^\text{SVD}=8$ (right panel) as compared to the fRG ground truth (left panel).
These results suggest also that the 4-point vertex function (in principle a complicated function of three variables) can indeed be very compactly parametrized by a small number of basis functions in each of the relevant two particle regimes. Whether this finding applies more generally to the vertex functions is an important open question.


{\it Outlook --} 
Our work presents an application of artificial intelligence to fRG, which successfully unveils a dimensionality reduced dynamics for the Hubbard model on a square lattice at specific sets of electron filling. Nonetheless, the relevance of the procedure outlined in our paper goes beyond the testbed cases considered. In particular, the identification of how to extract and manipulate relevant information encoded in the 4-point (or two-particle) vertex functions of many-electron problems, separating it from the non-relevant ones, represents a goal of utmost importance for the success of several cutting-edge quantum field theoretical methods for quantum materials. The 4-point vertex is, in fact, the building block of advanced approaches based on the solution of fRG flow equations~\cite{MetznerRMP} or resummation of parquet diagrams~\cite{Bickersbook2004}, including their most recent developments such as the multi-loop extension~\cite{Kugler2018,Kiese2021}  of fRG, the merger of fRG and DMFT~\cite{Taranto2014} and the diagrammatic extensions~\cite{Rohringer2018} of DMFT itself, specifically designed for computing the most challenging non-perturbative regimes.

In addition, the promising outcome of our deep learning-based fRG procedure naturally suggests its extension beyond the van Hove filling illustrated here as well as to more realistic models including non-local electronic interactions. It will be also important to explore whether transfer learning~\cite{tan2018survey} could  mitigate the burden of training deep nets for similar Hamiltonians on other lattice geometries, and whether a rich set of learned latent variables allows for extracting equations from the data (symbolic regression)~\cite{SINDYPNAS} and for a resolution increase of the vertex.

{\it Acknowledgments --} The authors are grateful to Ronny Thomale for providing the Fortran $N$-patch fRG code used to generate the ground truth data, and to Daniel Springer and Tobias M\"{u}ller for stimulating discussions. The research leading to these results has received funding from the European Union’s Horizon 2020 research and innovation programme under the Marie Sk{\l}odowska-Curie Grant Agreement No. 897276 (BITMAP). This work is funded by the Deutsche Forschungsgemeinschaft (DFG, German Research Foundation) through Project-ID 258499086 - SFB 1170 and through the W{\"u}rzburg-Dresden Cluster of Excellence on Complexity and Topology in Quantum Matter-ct.qmat Project- ID 390858490 - EXC 2147 as well as the Austrian Science Fund (FWF) through the project I 2794-N35. MM acknowledges support from the CCQ graduate fellowship in computational quantum physics. The Flatiron Institute is a division of the Simons Foundation.

\bibliography{biblio}

\begin{thebibliography}{79}
\expandafter\ifx\csname natexlab\endcsname\relax\def\natexlab#1{#1}\fi
\expandafter\ifx\csname bibnamefont\endcsname\relax
  \def\bibnamefont#1{#1}\fi
\expandafter\ifx\csname bibfnamefont\endcsname\relax
  \def\bibfnamefont#1{#1}\fi
\expandafter\ifx\csname citenamefont\endcsname\relax
  \def\citenamefont#1{#1}\fi
\expandafter\ifx\csname url\endcsname\relax
  \def\url#1{\texttt{#1}}\fi
\expandafter\ifx\csname urlprefix\endcsname\relax\def\urlprefix{URL }\fi
\providecommand{\bibinfo}[2]{#2}
\providecommand{\eprint}[2][]{\url{#2}}

\bibitem[{\citenamefont{Wilson and Kogut}(1974)}]{wilson1974renormalization}
\bibinfo{author}{\bibfnamefont{K.~G.} \bibnamefont{Wilson}} \bibnamefont{and}
  \bibinfo{author}{\bibfnamefont{J.~G.} \bibnamefont{Kogut}},
  \bibinfo{journal}{Phys. Rep.} \textbf{\bibinfo{volume}{12}},
  \bibinfo{pages}{75} (\bibinfo{year}{1974}).

\bibitem[{\citenamefont{Wilson}(1975)}]{WilsonRMP}
\bibinfo{author}{\bibfnamefont{K.~G.} \bibnamefont{Wilson}},
  \bibinfo{journal}{Rev. Mod. Phys.} \textbf{\bibinfo{volume}{47}},
  \bibinfo{pages}{773} (\bibinfo{year}{1975}).

\bibitem[{\citenamefont{Fisher}(1983)}]{Fisher1983}
\bibinfo{author}{\bibfnamefont{M.~E.} \bibnamefont{Fisher}},
  \emph{\bibinfo{title}{Scaling, Universality and Renormalization Group
  Theory}} (\bibinfo{publisher}{Springer, Berlin}, \bibinfo{year}{1983}).

\bibitem[{\citenamefont{Shankar}(1994)}]{ShankarRMP}
\bibinfo{author}{\bibfnamefont{R.}~\bibnamefont{Shankar}},
  \bibinfo{journal}{Rev. Mod. Phys.} \textbf{\bibinfo{volume}{66}},
  \bibinfo{pages}{129} (\bibinfo{year}{1994}).

\bibitem[{\citenamefont{Polchinski}(1992)}]{polchinski1992effective}
\bibinfo{author}{\bibfnamefont{J.}~\bibnamefont{Polchinski}}
  (\bibinfo{year}{1992}), \eprint{arXiv:hep-th/9210046}.

\bibitem[{\citenamefont{Wetterich}(1993)}]{Wetterich}
\bibinfo{author}{\bibfnamefont{C.}~\bibnamefont{Wetterich}},
  \bibinfo{journal}{Phys. Lett. B} \textbf{\bibinfo{volume}{301}},
  \bibinfo{pages}{90} (\bibinfo{year}{1993}).

\bibitem[{\citenamefont{Morris}(1994)}]{Morris}
\bibinfo{author}{\bibfnamefont{T.~R.} \bibnamefont{Morris}},
  \bibinfo{journal}{Int. J. Mod. Phys. A} \textbf{\bibinfo{volume}{9}},
  \bibinfo{pages}{2411} (\bibinfo{year}{1994}).

\bibitem[{\citenamefont{Salmhofer and Honerkamp}(2001)}]{SalmhoferPTP2001}
\bibinfo{author}{\bibfnamefont{M.}~\bibnamefont{Salmhofer}} \bibnamefont{and}
  \bibinfo{author}{\bibfnamefont{C.}~\bibnamefont{Honerkamp}},
  \bibinfo{journal}{Progr. Theoret. Phys.} \textbf{\bibinfo{volume}{105}},
  \bibinfo{pages}{1} (\bibinfo{year}{2001}).

\bibitem[{\citenamefont{Kopietz and Busche}(2001)}]{KopietzPRB2001}
\bibinfo{author}{\bibfnamefont{P.}~\bibnamefont{Kopietz}} \bibnamefont{and}
  \bibinfo{author}{\bibfnamefont{T.}~\bibnamefont{Busche}},
  \bibinfo{journal}{Phys. Rev. B} \textbf{\bibinfo{volume}{64}},
  \bibinfo{pages}{155101} (\bibinfo{year}{2001}).

\bibitem[{\citenamefont{Zanchi and Schulz}(2000)}]{ZanchiPRB2000}
\bibinfo{author}{\bibfnamefont{D.}~\bibnamefont{Zanchi}} \bibnamefont{and}
  \bibinfo{author}{\bibfnamefont{H.~J.} \bibnamefont{Schulz}},
  \bibinfo{journal}{Phys. Rev. B} \textbf{\bibinfo{volume}{61}},
  \bibinfo{pages}{13609} (\bibinfo{year}{2000}).

\bibitem[{\citenamefont{Halboth and
  Metzner}(2000{\natexlab{a}})}]{HalbothPRB2000}
\bibinfo{author}{\bibfnamefont{C.~J.} \bibnamefont{Halboth}} \bibnamefont{and}
  \bibinfo{author}{\bibfnamefont{W.}~\bibnamefont{Metzner}},
  \bibinfo{journal}{Phys. Rev. B} \textbf{\bibinfo{volume}{61}},
  \bibinfo{pages}{7364} (\bibinfo{year}{2000}{\natexlab{a}}).

\bibitem[{\citenamefont{Halboth and
  Metzner}(2000{\natexlab{b}})}]{HalbothPRL2000}
\bibinfo{author}{\bibfnamefont{C.~J.} \bibnamefont{Halboth}} \bibnamefont{and}
  \bibinfo{author}{\bibfnamefont{W.}~\bibnamefont{Metzner}},
  \bibinfo{journal}{Phys. Rev. Lett.} \textbf{\bibinfo{volume}{85}},
  \bibinfo{pages}{5162} (\bibinfo{year}{2000}{\natexlab{b}}).

\bibitem[{\citenamefont{Honerkamp et~al.}(2001)\citenamefont{Honerkamp,
  Salmhofer, Furukawa, and Rice}}]{HonerkampPRB2001}
\bibinfo{author}{\bibfnamefont{C.}~\bibnamefont{Honerkamp}},
  \bibinfo{author}{\bibfnamefont{M.}~\bibnamefont{Salmhofer}},
  \bibinfo{author}{\bibfnamefont{N.}~\bibnamefont{Furukawa}}, \bibnamefont{and}
  \bibinfo{author}{\bibfnamefont{T.~M.} \bibnamefont{Rice}},
  \bibinfo{journal}{Phys. Rev. B} \textbf{\bibinfo{volume}{63}},
  \bibinfo{pages}{035109} (\bibinfo{year}{2001}).

\bibitem[{\citenamefont{Honerkamp and
  Salmhofer}(2001{\natexlab{a}})}]{HonerkampSalmhoferPRL}
\bibinfo{author}{\bibfnamefont{C.}~\bibnamefont{Honerkamp}} \bibnamefont{and}
  \bibinfo{author}{\bibfnamefont{M.}~\bibnamefont{Salmhofer}},
  \bibinfo{journal}{Phys. Rev. Lett.} \textbf{\bibinfo{volume}{87}},
  \bibinfo{pages}{187004} (\bibinfo{year}{2001}{\natexlab{a}}).

\bibitem[{\citenamefont{Honerkamp and
  Salmhofer}(2001{\natexlab{b}})}]{HonerkampSalmhoferPRB}
\bibinfo{author}{\bibfnamefont{C.}~\bibnamefont{Honerkamp}} \bibnamefont{and}
  \bibinfo{author}{\bibfnamefont{M.}~\bibnamefont{Salmhofer}},
  \bibinfo{journal}{Phys. Rev. B} \textbf{\bibinfo{volume}{64}},
  \bibinfo{pages}{184516} (\bibinfo{year}{2001}{\natexlab{b}}).

\bibitem[{\citenamefont{Hille et~al.}(2020{\natexlab{a}})\citenamefont{Hille,
  Kugler, Eckhardt, He, Kauch, Honerkamp, Toschi, and Andergassen}}]{Hille2020}
\bibinfo{author}{\bibfnamefont{C.}~\bibnamefont{Hille}},
  \bibinfo{author}{\bibfnamefont{F.~B.} \bibnamefont{Kugler}},
  \bibinfo{author}{\bibfnamefont{C.~J.} \bibnamefont{Eckhardt}},
  \bibinfo{author}{\bibfnamefont{Y.-Y.} \bibnamefont{He}},
  \bibinfo{author}{\bibfnamefont{A.}~\bibnamefont{Kauch}},
  \bibinfo{author}{\bibfnamefont{C.}~\bibnamefont{Honerkamp}},
  \bibinfo{author}{\bibfnamefont{A.}~\bibnamefont{Toschi}}, \bibnamefont{and}
  \bibinfo{author}{\bibfnamefont{S.}~\bibnamefont{Andergassen}},
  \bibinfo{journal}{Phys. Rev. Research} \textbf{\bibinfo{volume}{2}},
  \bibinfo{pages}{033372} (\bibinfo{year}{2020}{\natexlab{a}}).

\bibitem[{\citenamefont{Hille et~al.}(2020{\natexlab{b}})\citenamefont{Hille,
  Rohe, Honerkamp, and Andergassen}}]{Hille2020B}
\bibinfo{author}{\bibfnamefont{C.}~\bibnamefont{Hille}},
  \bibinfo{author}{\bibfnamefont{D.}~\bibnamefont{Rohe}},
  \bibinfo{author}{\bibfnamefont{C.}~\bibnamefont{Honerkamp}},
  \bibnamefont{and}
  \bibinfo{author}{\bibfnamefont{S.}~\bibnamefont{Andergassen}},
  \bibinfo{journal}{Phys. Rev. Research} \textbf{\bibinfo{volume}{2}},
  \bibinfo{pages}{033068} (\bibinfo{year}{2020}{\natexlab{b}}).

\bibitem[{\citenamefont{Vilardi et~al.}(2020)\citenamefont{Vilardi, Bonetti,
  and Metzner}}]{Vilardi2020}
\bibinfo{author}{\bibfnamefont{D.}~\bibnamefont{Vilardi}},
  \bibinfo{author}{\bibfnamefont{P.~M.} \bibnamefont{Bonetti}},
  \bibnamefont{and} \bibinfo{author}{\bibfnamefont{W.}~\bibnamefont{Metzner}},
  \bibinfo{journal}{Phys. Rev. B} \textbf{\bibinfo{volume}{102}},
  \bibinfo{pages}{245128} (\bibinfo{year}{2020}).

\bibitem[{\citenamefont{Wang et~al.}(2009)\citenamefont{Wang, Zhai, Ran,
  Vishwanath, and Lee}}]{WangPRL2009}
\bibinfo{author}{\bibfnamefont{F.}~\bibnamefont{Wang}},
  \bibinfo{author}{\bibfnamefont{H.}~\bibnamefont{Zhai}},
  \bibinfo{author}{\bibfnamefont{Y.}~\bibnamefont{Ran}},
  \bibinfo{author}{\bibfnamefont{A.}~\bibnamefont{Vishwanath}},
  \bibnamefont{and} \bibinfo{author}{\bibfnamefont{D.-H.} \bibnamefont{Lee}},
  \bibinfo{journal}{Phys. Rev. Lett.} \textbf{\bibinfo{volume}{102}},
  \bibinfo{pages}{047005} (\bibinfo{year}{2009}).

\bibitem[{\citenamefont{Thomale et~al.}(2009)\citenamefont{Thomale, Platt, Hu,
  Honerkamp, and Bernevig}}]{ThomalePRB2009}
\bibinfo{author}{\bibfnamefont{R.}~\bibnamefont{Thomale}},
  \bibinfo{author}{\bibfnamefont{C.}~\bibnamefont{Platt}},
  \bibinfo{author}{\bibfnamefont{J.}~\bibnamefont{Hu}},
  \bibinfo{author}{\bibfnamefont{C.}~\bibnamefont{Honerkamp}},
  \bibnamefont{and} \bibinfo{author}{\bibfnamefont{B.~A.}
  \bibnamefont{Bernevig}}, \bibinfo{journal}{Phys. Rev. B}
  \textbf{\bibinfo{volume}{80}}, \bibinfo{pages}{180505}
  (\bibinfo{year}{2009}).

\bibitem[{\citenamefont{Thomale
  et~al.}(2011{\natexlab{a}})\citenamefont{Thomale, Platt, Hanke, and
  Bernevig}}]{ThomalePRL2011a}
\bibinfo{author}{\bibfnamefont{R.}~\bibnamefont{Thomale}},
  \bibinfo{author}{\bibfnamefont{C.}~\bibnamefont{Platt}},
  \bibinfo{author}{\bibfnamefont{W.}~\bibnamefont{Hanke}}, \bibnamefont{and}
  \bibinfo{author}{\bibfnamefont{B.~A.} \bibnamefont{Bernevig}},
  \bibinfo{journal}{Phys. Rev. Lett.} \textbf{\bibinfo{volume}{106}},
  \bibinfo{pages}{187003} (\bibinfo{year}{2011}{\natexlab{a}}).

\bibitem[{\citenamefont{Thomale
  et~al.}(2011{\natexlab{b}})\citenamefont{Thomale, Platt, Hanke, Hu, and
  Bernevig}}]{ThomalePRL2011b}
\bibinfo{author}{\bibfnamefont{R.}~\bibnamefont{Thomale}},
  \bibinfo{author}{\bibfnamefont{C.}~\bibnamefont{Platt}},
  \bibinfo{author}{\bibfnamefont{W.}~\bibnamefont{Hanke}},
  \bibinfo{author}{\bibfnamefont{J.}~\bibnamefont{Hu}}, \bibnamefont{and}
  \bibinfo{author}{\bibfnamefont{B.~A.} \bibnamefont{Bernevig}},
  \bibinfo{journal}{Phys. Rev. Lett.} \textbf{\bibinfo{volume}{107}},
  \bibinfo{pages}{117001} (\bibinfo{year}{2011}{\natexlab{b}}).

\bibitem[{\citenamefont{Kiesel et~al.}(2013)\citenamefont{Kiesel, Platt, Hanke,
  and Thomale}}]{KieselPRL2013}
\bibinfo{author}{\bibfnamefont{M.~L.} \bibnamefont{Kiesel}},
  \bibinfo{author}{\bibfnamefont{C.}~\bibnamefont{Platt}},
  \bibinfo{author}{\bibfnamefont{W.}~\bibnamefont{Hanke}}, \bibnamefont{and}
  \bibinfo{author}{\bibfnamefont{R.}~\bibnamefont{Thomale}},
  \bibinfo{journal}{Phys. Rev. Lett.} \textbf{\bibinfo{volume}{111}},
  \bibinfo{pages}{097001} (\bibinfo{year}{2013}).

\bibitem[{\citenamefont{Honerkamp}(2008)}]{HonerkampPRL2008}
\bibinfo{author}{\bibfnamefont{C.}~\bibnamefont{Honerkamp}},
  \bibinfo{journal}{Phys. Rev. Lett.} \textbf{\bibinfo{volume}{100}},
  \bibinfo{pages}{146404} (\bibinfo{year}{2008}).

\bibitem[{\citenamefont{Kiesel et~al.}(2012)\citenamefont{Kiesel, Platt, Hanke,
  Abanin, and Thomale}}]{KieselPRB2012}
\bibinfo{author}{\bibfnamefont{M.~L.} \bibnamefont{Kiesel}},
  \bibinfo{author}{\bibfnamefont{C.}~\bibnamefont{Platt}},
  \bibinfo{author}{\bibfnamefont{W.}~\bibnamefont{Hanke}},
  \bibinfo{author}{\bibfnamefont{D.~A.} \bibnamefont{Abanin}},
  \bibnamefont{and} \bibinfo{author}{\bibfnamefont{R.}~\bibnamefont{Thomale}},
  \bibinfo{journal}{Phys. Rev. B} \textbf{\bibinfo{volume}{86}},
  \bibinfo{pages}{020507} (\bibinfo{year}{2012}).

\bibitem[{\citenamefont{Kennes et~al.}(2018)\citenamefont{Kennes, Lischner, and
  Karrasch}}]{KennesPRB2018}
\bibinfo{author}{\bibfnamefont{D.~M.} \bibnamefont{Kennes}},
  \bibinfo{author}{\bibfnamefont{J.}~\bibnamefont{Lischner}}, \bibnamefont{and}
  \bibinfo{author}{\bibfnamefont{C.}~\bibnamefont{Karrasch}},
  \bibinfo{journal}{Phys. Rev. B} \textbf{\bibinfo{volume}{98}},
  \bibinfo{pages}{241407} (\bibinfo{year}{2018}).

\bibitem[{\citenamefont{Di~Sante et~al.}(2019)\citenamefont{Di~Sante, Wu, Fink,
  Hanke, and Thomale}}]{GeMoS2fRG}
\bibinfo{author}{\bibfnamefont{D.}~\bibnamefont{Di~Sante}},
  \bibinfo{author}{\bibfnamefont{X.}~\bibnamefont{Wu}},
  \bibinfo{author}{\bibfnamefont{M.}~\bibnamefont{Fink}},
  \bibinfo{author}{\bibfnamefont{W.}~\bibnamefont{Hanke}}, \bibnamefont{and}
  \bibinfo{author}{\bibfnamefont{R.}~\bibnamefont{Thomale}},
  \bibinfo{journal}{Phys. Rev. B} \textbf{\bibinfo{volume}{99}},
  \bibinfo{pages}{201106} (\bibinfo{year}{2019}).

\bibitem[{\citenamefont{Wu et~al.}(2019)\citenamefont{Wu, Fink, Hanke, Thomale,
  and Di~Sante}}]{WuPRB2019}
\bibinfo{author}{\bibfnamefont{X.}~\bibnamefont{Wu}},
  \bibinfo{author}{\bibfnamefont{M.}~\bibnamefont{Fink}},
  \bibinfo{author}{\bibfnamefont{W.}~\bibnamefont{Hanke}},
  \bibinfo{author}{\bibfnamefont{R.}~\bibnamefont{Thomale}}, \bibnamefont{and}
  \bibinfo{author}{\bibfnamefont{D.}~\bibnamefont{Di~Sante}},
  \bibinfo{journal}{Phys. Rev. B} \textbf{\bibinfo{volume}{100}},
  \bibinfo{pages}{041117} (\bibinfo{year}{2019}).

\bibitem[{\citenamefont{Chillal et~al.}(2020)\citenamefont{Chillal, Iqbal,
  Jeschke, Rodriguez-Rivera, Bewley, Manuel, Khalyavin, Steffens, Thomale,
  Islam et~al.}}]{Chillal2020}
\bibinfo{author}{\bibfnamefont{S.}~\bibnamefont{Chillal}},
  \bibinfo{author}{\bibfnamefont{Y.}~\bibnamefont{Iqbal}},
  \bibinfo{author}{\bibfnamefont{H.~O.} \bibnamefont{Jeschke}},
  \bibinfo{author}{\bibfnamefont{J.~A.} \bibnamefont{Rodriguez-Rivera}},
  \bibinfo{author}{\bibfnamefont{R.}~\bibnamefont{Bewley}},
  \bibinfo{author}{\bibfnamefont{P.}~\bibnamefont{Manuel}},
  \bibinfo{author}{\bibfnamefont{D.}~\bibnamefont{Khalyavin}},
  \bibinfo{author}{\bibfnamefont{P.}~\bibnamefont{Steffens}},
  \bibinfo{author}{\bibfnamefont{R.}~\bibnamefont{Thomale}},
  \bibinfo{author}{\bibfnamefont{A.~T. M.~N.} \bibnamefont{Islam}},
  \bibnamefont{et~al.}, \bibinfo{journal}{Nature Communications}
  \textbf{\bibinfo{volume}{11}}, \bibinfo{pages}{2348} (\bibinfo{year}{2020}).

\bibitem[{\citenamefont{Buessen et~al.}(2018)\citenamefont{Buessen, Hering,
  Reuther, and Trebst}}]{BuessenPRL2018}
\bibinfo{author}{\bibfnamefont{F.~L.} \bibnamefont{Buessen}},
  \bibinfo{author}{\bibfnamefont{M.}~\bibnamefont{Hering}},
  \bibinfo{author}{\bibfnamefont{J.}~\bibnamefont{Reuther}}, \bibnamefont{and}
  \bibinfo{author}{\bibfnamefont{S.}~\bibnamefont{Trebst}},
  \bibinfo{journal}{Phys. Rev. Lett.} \textbf{\bibinfo{volume}{120}},
  \bibinfo{pages}{057201} (\bibinfo{year}{2018}).

\bibitem[{\citenamefont{Wentzell et~al.}(2020)\citenamefont{Wentzell, Li,
  Tagliavini, Taranto, Rohringer, Held, Toschi, and Andergassen}}]{NilsPRB2020}
\bibinfo{author}{\bibfnamefont{N.}~\bibnamefont{Wentzell}},
  \bibinfo{author}{\bibfnamefont{G.}~\bibnamefont{Li}},
  \bibinfo{author}{\bibfnamefont{A.}~\bibnamefont{Tagliavini}},
  \bibinfo{author}{\bibfnamefont{C.}~\bibnamefont{Taranto}},
  \bibinfo{author}{\bibfnamefont{G.}~\bibnamefont{Rohringer}},
  \bibinfo{author}{\bibfnamefont{K.}~\bibnamefont{Held}},
  \bibinfo{author}{\bibfnamefont{A.}~\bibnamefont{Toschi}}, \bibnamefont{and}
  \bibinfo{author}{\bibfnamefont{S.}~\bibnamefont{Andergassen}},
  \bibinfo{journal}{Phys. Rev. B} \textbf{\bibinfo{volume}{102}},
  \bibinfo{pages}{085106} (\bibinfo{year}{2020}).

\bibitem[{\citenamefont{Kaye et~al.}(2022)\citenamefont{Kaye, Chen, and
  Parcollet}}]{Parcollet}
\bibinfo{author}{\bibfnamefont{J.}~\bibnamefont{Kaye}},
  \bibinfo{author}{\bibfnamefont{K.}~\bibnamefont{Chen}}, \bibnamefont{and}
  \bibinfo{author}{\bibfnamefont{O.}~\bibnamefont{Parcollet}},
  \bibinfo{journal}{Phys. Rev. B} \textbf{\bibinfo{volume}{105}},
  \bibinfo{pages}{235115} (\bibinfo{year}{2022}).

\bibitem[{\citenamefont{Husemann and Salmhofer}(2009)}]{HusemannPRB2009}
\bibinfo{author}{\bibfnamefont{C.}~\bibnamefont{Husemann}} \bibnamefont{and}
  \bibinfo{author}{\bibfnamefont{M.}~\bibnamefont{Salmhofer}},
  \bibinfo{journal}{Phys. Rev. B} \textbf{\bibinfo{volume}{79}},
  \bibinfo{pages}{195125} (\bibinfo{year}{2009}).

\bibitem[{\citenamefont{Wang et~al.}(2012)\citenamefont{Wang, Xiang, Wang,
  Wang, Yang, and Lee}}]{WangPRB2012}
\bibinfo{author}{\bibfnamefont{W.-S.} \bibnamefont{Wang}},
  \bibinfo{author}{\bibfnamefont{Y.-Y.} \bibnamefont{Xiang}},
  \bibinfo{author}{\bibfnamefont{Q.-H.} \bibnamefont{Wang}},
  \bibinfo{author}{\bibfnamefont{F.}~\bibnamefont{Wang}},
  \bibinfo{author}{\bibfnamefont{F.}~\bibnamefont{Yang}}, \bibnamefont{and}
  \bibinfo{author}{\bibfnamefont{D.-H.} \bibnamefont{Lee}},
  \bibinfo{journal}{Phys. Rev. B} \textbf{\bibinfo{volume}{85}},
  \bibinfo{pages}{035414} (\bibinfo{year}{2012}).

\bibitem[{\citenamefont{Lichtenstein et~al.}(2017)\citenamefont{Lichtenstein,
  {Sánchez de la Peña}, Rohe, {Di Napoli}, Honerkamp, and
  Maier}}]{LichtensteinTUfRG}
\bibinfo{author}{\bibfnamefont{J.}~\bibnamefont{Lichtenstein}},
  \bibinfo{author}{\bibfnamefont{D.}~\bibnamefont{{Sánchez de la Peña}}},
  \bibinfo{author}{\bibfnamefont{D.}~\bibnamefont{Rohe}},
  \bibinfo{author}{\bibfnamefont{E.}~\bibnamefont{{Di Napoli}}},
  \bibinfo{author}{\bibfnamefont{C.}~\bibnamefont{Honerkamp}},
  \bibnamefont{and} \bibinfo{author}{\bibfnamefont{S.}~\bibnamefont{Maier}},
  \bibinfo{journal}{Computer Physics Communications}
  \textbf{\bibinfo{volume}{213}}, \bibinfo{pages}{100} (\bibinfo{year}{2017}).

\bibitem[{\citenamefont{Carleo et~al.}(2019)\citenamefont{Carleo, Cirac,
  Cranmer, Daudet, Schuld, Tishby, Vogt-Maranto, and Zdeborov\'a}}]{CarleoRMP}
\bibinfo{author}{\bibfnamefont{G.}~\bibnamefont{Carleo}},
  \bibinfo{author}{\bibfnamefont{I.}~\bibnamefont{Cirac}},
  \bibinfo{author}{\bibfnamefont{K.}~\bibnamefont{Cranmer}},
  \bibinfo{author}{\bibfnamefont{L.}~\bibnamefont{Daudet}},
  \bibinfo{author}{\bibfnamefont{M.}~\bibnamefont{Schuld}},
  \bibinfo{author}{\bibfnamefont{N.}~\bibnamefont{Tishby}},
  \bibinfo{author}{\bibfnamefont{L.}~\bibnamefont{Vogt-Maranto}},
  \bibnamefont{and}
  \bibinfo{author}{\bibfnamefont{L.}~\bibnamefont{Zdeborov\'a}},
  \bibinfo{journal}{Rev. Mod. Phys.} \textbf{\bibinfo{volume}{91}},
  \bibinfo{pages}{045002} (\bibinfo{year}{2019}).

\bibitem[{\citenamefont{Snyder et~al.}(2012)\citenamefont{Snyder, Rupp, Hansen,
  M\"uller, and Burke}}]{BurkePRL2012}
\bibinfo{author}{\bibfnamefont{J.~C.} \bibnamefont{Snyder}},
  \bibinfo{author}{\bibfnamefont{M.}~\bibnamefont{Rupp}},
  \bibinfo{author}{\bibfnamefont{K.}~\bibnamefont{Hansen}},
  \bibinfo{author}{\bibfnamefont{K.-R.} \bibnamefont{M\"uller}},
  \bibnamefont{and} \bibinfo{author}{\bibfnamefont{K.}~\bibnamefont{Burke}},
  \bibinfo{journal}{Phys. Rev. Lett.} \textbf{\bibinfo{volume}{108}},
  \bibinfo{pages}{253002} (\bibinfo{year}{2012}).

\bibitem[{\citenamefont{Brockherde et~al.}(2017)\citenamefont{Brockherde, Vogt,
  Li, Tuckerman, Burke, and M{\"u}ller}}]{BurkeNatComm2017}
\bibinfo{author}{\bibfnamefont{F.}~\bibnamefont{Brockherde}},
  \bibinfo{author}{\bibfnamefont{L.}~\bibnamefont{Vogt}},
  \bibinfo{author}{\bibfnamefont{L.}~\bibnamefont{Li}},
  \bibinfo{author}{\bibfnamefont{M.~E.} \bibnamefont{Tuckerman}},
  \bibinfo{author}{\bibfnamefont{K.}~\bibnamefont{Burke}}, \bibnamefont{and}
  \bibinfo{author}{\bibfnamefont{K.-R.} \bibnamefont{M{\"u}ller}},
  \bibinfo{journal}{Nat. Commun.} \textbf{\bibinfo{volume}{8}},
  \bibinfo{pages}{872} (\bibinfo{year}{2017}).

\bibitem[{\citenamefont{Moreno et~al.}(2020)\citenamefont{Moreno, Carleo, and
  Georges}}]{GeorgesPRL}
\bibinfo{author}{\bibfnamefont{J.~R.} \bibnamefont{Moreno}},
  \bibinfo{author}{\bibfnamefont{G.}~\bibnamefont{Carleo}}, \bibnamefont{and}
  \bibinfo{author}{\bibfnamefont{A.}~\bibnamefont{Georges}},
  \bibinfo{journal}{Phys. Rev. Lett.} \textbf{\bibinfo{volume}{125}},
  \bibinfo{pages}{076402} (\bibinfo{year}{2020}).

\bibitem[{\citenamefont{Kirkpatrick et~al.}(2021)\citenamefont{Kirkpatrick,
  McMorrow, Turban, Gaunt, Spencer, Matthews, Obika, Thiry, Fortunato, Pfau
  et~al.}}]{KirkpatrickScience2021}
\bibinfo{author}{\bibfnamefont{J.}~\bibnamefont{Kirkpatrick}},
  \bibinfo{author}{\bibfnamefont{B.}~\bibnamefont{McMorrow}},
  \bibinfo{author}{\bibfnamefont{D.~H.~P.} \bibnamefont{Turban}},
  \bibinfo{author}{\bibfnamefont{A.~L.} \bibnamefont{Gaunt}},
  \bibinfo{author}{\bibfnamefont{J.~S.} \bibnamefont{Spencer}},
  \bibinfo{author}{\bibfnamefont{A.~G. D.~G.} \bibnamefont{Matthews}},
  \bibinfo{author}{\bibfnamefont{A.}~\bibnamefont{Obika}},
  \bibinfo{author}{\bibfnamefont{L.}~\bibnamefont{Thiry}},
  \bibinfo{author}{\bibfnamefont{M.}~\bibnamefont{Fortunato}},
  \bibinfo{author}{\bibfnamefont{D.}~\bibnamefont{Pfau}}, \bibnamefont{et~al.},
  \bibinfo{journal}{Science} \textbf{\bibinfo{volume}{374}},
  \bibinfo{pages}{1385} (\bibinfo{year}{2021}).

\bibitem[{\citenamefont{Arsenault et~al.}(2014)\citenamefont{Arsenault,
  Lopez-Bezanilla, von Lilienfeld, and Millis}}]{MillisPRB2014}
\bibinfo{author}{\bibfnamefont{L.}~\bibnamefont{Arsenault}},
  \bibinfo{author}{\bibfnamefont{A.}~\bibnamefont{Lopez-Bezanilla}},
  \bibinfo{author}{\bibfnamefont{O.~A.} \bibnamefont{von Lilienfeld}},
  \bibnamefont{and} \bibinfo{author}{\bibfnamefont{A.~J.}
  \bibnamefont{Millis}}, \bibinfo{journal}{Phys. Rev. B}
  \textbf{\bibinfo{volume}{90}}, \bibinfo{pages}{155136}
  (\bibinfo{year}{2014}).

\bibitem[{\citenamefont{Rogers et~al.}(2021)\citenamefont{Rogers, Lee, Pakdel,
  Xu, Dobrosavljevi\ifmmode~\acute{c}\else \'{c}\fi{}, Yao, Christiansen, and
  Lanat\`a}}]{LanataPRR2021}
\bibinfo{author}{\bibfnamefont{J.}~\bibnamefont{Rogers}},
  \bibinfo{author}{\bibfnamefont{T.-H.} \bibnamefont{Lee}},
  \bibinfo{author}{\bibfnamefont{S.}~\bibnamefont{Pakdel}},
  \bibinfo{author}{\bibfnamefont{W.}~\bibnamefont{Xu}},
  \bibinfo{author}{\bibfnamefont{V.}~\bibnamefont{Dobrosavljevi\ifmmode~\acute{c}\else
  \'{c}\fi{}}}, \bibinfo{author}{\bibfnamefont{Y.-X.} \bibnamefont{Yao}},
  \bibinfo{author}{\bibfnamefont{O.}~\bibnamefont{Christiansen}},
  \bibnamefont{and} \bibinfo{author}{\bibfnamefont{N.}~\bibnamefont{Lanat\`a}},
  \bibinfo{journal}{Phys. Rev. Research} \textbf{\bibinfo{volume}{3}},
  \bibinfo{pages}{013101} (\bibinfo{year}{2021}).

\bibitem[{\citenamefont{Song and Lee}(2019)}]{LeeDMFT2019}
\bibinfo{author}{\bibfnamefont{T.}~\bibnamefont{Song}} \bibnamefont{and}
  \bibinfo{author}{\bibfnamefont{H.}~\bibnamefont{Lee}},
  \bibinfo{journal}{Phys. Rev. B} \textbf{\bibinfo{volume}{100}},
  \bibinfo{pages}{045153} (\bibinfo{year}{2019}).

\bibitem[{\citenamefont{Sheridan et~al.}(2021)\citenamefont{Sheridan, Rhodes,
  Jamet, Rungger, and Weber}}]{sheridan2021datadriven}
\bibinfo{author}{\bibfnamefont{E.}~\bibnamefont{Sheridan}},
  \bibinfo{author}{\bibfnamefont{C.}~\bibnamefont{Rhodes}},
  \bibinfo{author}{\bibfnamefont{F.}~\bibnamefont{Jamet}},
  \bibinfo{author}{\bibfnamefont{I.}~\bibnamefont{Rungger}}, \bibnamefont{and}
  \bibinfo{author}{\bibfnamefont{C.}~\bibnamefont{Weber}},
  \bibinfo{journal}{Phys. Rev. B} \textbf{\bibinfo{volume}{104}},
  \bibinfo{pages}{205120} (\bibinfo{year}{2021}).

\bibitem[{\citenamefont{Li and Wang}(2018)}]{LeiWang}
\bibinfo{author}{\bibfnamefont{S.-H.} \bibnamefont{Li}} \bibnamefont{and}
  \bibinfo{author}{\bibfnamefont{L.}~\bibnamefont{Wang}},
  \bibinfo{journal}{Phys. Rev. Lett.} \textbf{\bibinfo{volume}{121}},
  \bibinfo{pages}{260601} (\bibinfo{year}{2018}).

\bibitem[{\citenamefont{Rigo and Mitchell}(2022)}]{RigoNRG}
\bibinfo{author}{\bibfnamefont{J.~B.} \bibnamefont{Rigo}} \bibnamefont{and}
  \bibinfo{author}{\bibfnamefont{A.~K.} \bibnamefont{Mitchell}},
  \bibinfo{journal}{Phys. Rev. Research} \textbf{\bibinfo{volume}{4}},
  \bibinfo{pages}{013227} (\bibinfo{year}{2022}).

\bibitem[{\citenamefont{Carleo and Troyer}(2017)}]{CarleoScience}
\bibinfo{author}{\bibfnamefont{G.}~\bibnamefont{Carleo}} \bibnamefont{and}
  \bibinfo{author}{\bibfnamefont{M.}~\bibnamefont{Troyer}},
  \bibinfo{journal}{Science} \textbf{\bibinfo{volume}{355}},
  \bibinfo{pages}{602} (\bibinfo{year}{2017}).

\bibitem[{\citenamefont{Chen et~al.}(2018)\citenamefont{Chen, Rubanova,
  Bettencourt, and Duvenaud}}]{Chen2018}
\bibinfo{author}{\bibfnamefont{R.~T.~Q.} \bibnamefont{Chen}},
  \bibinfo{author}{\bibfnamefont{Y.}~\bibnamefont{Rubanova}},
  \bibinfo{author}{\bibfnamefont{J.}~\bibnamefont{Bettencourt}},
  \bibnamefont{and} \bibinfo{author}{\bibfnamefont{D.}~\bibnamefont{Duvenaud}},
  \bibinfo{journal}{NIPs} \textbf{\bibinfo{volume}{109}}, \bibinfo{pages}{31}
  (\bibinfo{year}{2018}).

\bibitem[{\citenamefont{Schmid}(2010)}]{schmid_dynamic_2010}
\bibinfo{author}{\bibfnamefont{P.~J.} \bibnamefont{Schmid}},
  \bibinfo{journal}{Journal of Fluid Mechanics} \textbf{\bibinfo{volume}{656}},
  \bibinfo{pages}{5} (\bibinfo{year}{2010}).

\bibitem[{\citenamefont{Rowley et~al.}(2009)\citenamefont{Rowley, Mezi{\'c},
  Bagheri, Schlatter, and Henningson}}]{rowley2009spectral}
\bibinfo{author}{\bibfnamefont{C.~W.} \bibnamefont{Rowley}},
  \bibinfo{author}{\bibfnamefont{I.}~\bibnamefont{Mezi{\'c}}},
  \bibinfo{author}{\bibfnamefont{S.}~\bibnamefont{Bagheri}},
  \bibinfo{author}{\bibfnamefont{P.}~\bibnamefont{Schlatter}},
  \bibnamefont{and} \bibinfo{author}{\bibfnamefont{D.~S.}
  \bibnamefont{Henningson}}, \bibinfo{journal}{Journal of Fluid Mechanics}
  \textbf{\bibinfo{volume}{641}}, \bibinfo{pages}{115} (\bibinfo{year}{2009}).

\bibitem[{\citenamefont{Quarteroni and Rozza}(2014)}]{Quarteroni}
\bibinfo{editor}{\bibfnamefont{A.}~\bibnamefont{Quarteroni}} \bibnamefont{and}
  \bibinfo{editor}{\bibfnamefont{G.}~\bibnamefont{Rozza}}, eds.,
  \emph{\bibinfo{title}{Reduced Order Methods for Modeling and Computational
  Reduction}} (\bibinfo{publisher}{Springer, Cham}, \bibinfo{year}{2014}).

\bibitem[{\citenamefont{Metzner et~al.}(2012)\citenamefont{Metzner, Salmhofer,
  Honerkamp, Meden, and Sch\"onhammer}}]{MetznerRMP}
\bibinfo{author}{\bibfnamefont{W.}~\bibnamefont{Metzner}},
  \bibinfo{author}{\bibfnamefont{M.}~\bibnamefont{Salmhofer}},
  \bibinfo{author}{\bibfnamefont{C.}~\bibnamefont{Honerkamp}},
  \bibinfo{author}{\bibfnamefont{V.}~\bibnamefont{Meden}}, \bibnamefont{and}
  \bibinfo{author}{\bibfnamefont{K.}~\bibnamefont{Sch\"onhammer}},
  \bibinfo{journal}{Rev. Mod. Phys.} \textbf{\bibinfo{volume}{84}},
  \bibinfo{pages}{299} (\bibinfo{year}{2012}).

\bibitem[{\citenamefont{Platt et~al.}(2013)\citenamefont{Platt, Hanke, and
  Thomale}}]{PlattReview}
\bibinfo{author}{\bibfnamefont{C.}~\bibnamefont{Platt}},
  \bibinfo{author}{\bibfnamefont{W.}~\bibnamefont{Hanke}}, \bibnamefont{and}
  \bibinfo{author}{\bibfnamefont{R.}~\bibnamefont{Thomale}},
  \bibinfo{journal}{Advances in Physics} \textbf{\bibinfo{volume}{62}},
  \bibinfo{pages}{453} (\bibinfo{year}{2013}).

\bibitem[{\citenamefont{Kopietz et~al.}(2010)\citenamefont{Kopietz, Bartosch,
  and Sch\"utz}}]{KopietzBook}
\bibinfo{author}{\bibfnamefont{P.}~\bibnamefont{Kopietz}},
  \bibinfo{author}{\bibfnamefont{L.}~\bibnamefont{Bartosch}}, \bibnamefont{and}
  \bibinfo{author}{\bibfnamefont{F.}~\bibnamefont{Sch\"utz}},
  \emph{\bibinfo{title}{Introduction to the Functional Renormalization Group}}
  (\bibinfo{publisher}{Springer-Verlag, Berlin Heidelberg},
  \bibinfo{year}{2010}).

\bibitem[{not({\natexlab{a}})}]{note1}
\bibinfo{note}{Commonly, the feedback of the 2-point vertex function
  $\Sigma^\Lambda(\text{k})$ in $\text{L}^\Lambda$ is
  neglected~\cite{MetznerRMP,PlattReview}.}

\bibitem[{not({\natexlab{b}})}]{note4}
\bibinfo{note}{The reader interested in how the contributions to the effective
  action are classified as relevant, marginal and irrelevant, within the fRG
  framework, is referred to Refs.~\cite{KopietzPRB2001,MetznerRMP,KopietzBook},
  where rescaled versions of the fRG flow equations are derived and couplings
  are classified.}

\bibitem[{Sup()}]{SuppMatt}
\bibinfo{note}{The Supplementary Material at https://... contains further
  details and results. It also includes
  Refs.~\cite{Paszke2019,NumPy,SciPy,Matplotlib,sklearn}.}

\bibitem[{\citenamefont{Goodfellow et~al.}(2016)\citenamefont{Goodfellow,
  Bengio, and Courville}}]{goodfellow2016deep}
\bibinfo{author}{\bibfnamefont{I.}~\bibnamefont{Goodfellow}},
  \bibinfo{author}{\bibfnamefont{Y.}~\bibnamefont{Bengio}}, \bibnamefont{and}
  \bibinfo{author}{\bibfnamefont{A.}~\bibnamefont{Courville}},
  \emph{\bibinfo{title}{Deep learning}} (\bibinfo{publisher}{MIT press},
  \bibinfo{year}{2016}).

\bibitem[{\citenamefont{Vaswani et~al.}(2017)\citenamefont{Vaswani, Shazeer,
  Parmar, Uszkoreit, Jones, Gomez, Kaiser, and
  Polosukhin}}]{vaswani2017attention}
\bibinfo{author}{\bibfnamefont{A.}~\bibnamefont{Vaswani}},
  \bibinfo{author}{\bibfnamefont{N.}~\bibnamefont{Shazeer}},
  \bibinfo{author}{\bibfnamefont{N.}~\bibnamefont{Parmar}},
  \bibinfo{author}{\bibfnamefont{J.}~\bibnamefont{Uszkoreit}},
  \bibinfo{author}{\bibfnamefont{L.}~\bibnamefont{Jones}},
  \bibinfo{author}{\bibfnamefont{A.~N.} \bibnamefont{Gomez}},
  \bibinfo{author}{\bibfnamefont{{\L}.}~\bibnamefont{Kaiser}},
  \bibnamefont{and}
  \bibinfo{author}{\bibfnamefont{I.}~\bibnamefont{Polosukhin}},
  \bibinfo{journal}{Advances in neural information processing systems}
  \textbf{\bibinfo{volume}{30}} (\bibinfo{year}{2017}).

\bibitem[{\citenamefont{Lee and Parish}(2021)}]{Lee2021}
\bibinfo{author}{\bibfnamefont{K.}~\bibnamefont{Lee}} \bibnamefont{and}
  \bibinfo{author}{\bibfnamefont{E.~J.} \bibnamefont{Parish}},
  \bibinfo{journal}{Proc. R. Soc. A Math. Phys. Eng. Sci.}
  \textbf{\bibinfo{volume}{477}}, \bibinfo{pages}{20210162}
  (\bibinfo{year}{2021}).

\bibitem[{Neu()}]{NeuralFRG}
\bibinfo{note}{\lowercase{h}ttps://github.com/BITMAPdds/NeuralFRG}.

\bibitem[{\citenamefont{Kingma and Ba}(2015)}]{Kingma2015}
\bibinfo{author}{\bibfnamefont{D.~P.} \bibnamefont{Kingma}} \bibnamefont{and}
  \bibinfo{author}{\bibfnamefont{J.~L.} \bibnamefont{Ba}}, in
  \emph{\bibinfo{booktitle}{3rd Int. Conf. Learn. Represent. ICLR 2015 - Conf.
  Track Proc.}} (\bibinfo{publisher}{International Conference on Learning
  Representations, ICLR}, \bibinfo{year}{2015}), \eprint{1412.6980}.

\bibitem[{\citenamefont{Lin and Hirsch}(1987)}]{LinPRB1987}
\bibinfo{author}{\bibfnamefont{H.~Q.} \bibnamefont{Lin}} \bibnamefont{and}
  \bibinfo{author}{\bibfnamefont{J.~E.} \bibnamefont{Hirsch}},
  \bibinfo{journal}{Phys. Rev. B} \textbf{\bibinfo{volume}{35}},
  \bibinfo{pages}{3359} (\bibinfo{year}{1987}).

\bibitem[{\citenamefont{Lloyd}(1982)}]{Kmeans}
\bibinfo{author}{\bibfnamefont{S.}~\bibnamefont{Lloyd}}, \bibinfo{journal}{IEEE
  Transactions on Information Theory} \textbf{\bibinfo{volume}{28}},
  \bibinfo{pages}{129} (\bibinfo{year}{1982}).

\bibitem[{\citenamefont{Williams et~al.}(2014)\citenamefont{Williams, Rowley,
  and Kevrekidis}}]{williams2014kernel}
\bibinfo{author}{\bibfnamefont{M.~O.} \bibnamefont{Williams}},
  \bibinfo{author}{\bibfnamefont{C.~W.} \bibnamefont{Rowley}},
  \bibnamefont{and} \bibinfo{author}{\bibfnamefont{I.~G.}
  \bibnamefont{Kevrekidis}} (\bibinfo{year}{2014}), \eprint{arXiv:1411.2260}.

\bibitem[{\citenamefont{Kutz et~al.}(2016)\citenamefont{Kutz, Brunton, Brunton,
  and Proctor}}]{KutzBook}
\bibinfo{author}{\bibfnamefont{J.~N.} \bibnamefont{Kutz}},
  \bibinfo{author}{\bibfnamefont{S.~L.} \bibnamefont{Brunton}},
  \bibinfo{author}{\bibfnamefont{B.~W.} \bibnamefont{Brunton}},
  \bibnamefont{and} \bibinfo{author}{\bibfnamefont{J.~L.}
  \bibnamefont{Proctor}}, \emph{\bibinfo{title}{Dynamic Mode Decomposition:
  Data-Driven Modeling of Complex Systems}} (\bibinfo{publisher}{Society for
  Industrial and Applied Mathematics, Philadelphia}, \bibinfo{year}{2016}).

\bibitem[{\citenamefont{Jovanović et~al.}(2014)\citenamefont{Jovanović,
  Schmid, and Nichols}}]{Jovanovic2014}
\bibinfo{author}{\bibfnamefont{M.~R.} \bibnamefont{Jovanović}},
  \bibinfo{author}{\bibfnamefont{P.~J.} \bibnamefont{Schmid}},
  \bibnamefont{and} \bibinfo{author}{\bibfnamefont{J.~W.}
  \bibnamefont{Nichols}}, \bibinfo{journal}{Physics of Fluids}
  \textbf{\bibinfo{volume}{26}}, \bibinfo{pages}{024103}
  (\bibinfo{year}{2014}).

\bibitem[{\citenamefont{Bickers}(2004)}]{Bickersbook2004}
\bibinfo{author}{\bibfnamefont{N.~E.} \bibnamefont{Bickers}}, in
  \emph{\bibinfo{booktitle}{Theoretical Methods for Strongly Correlated
  Electrons}}, edited by
  \bibinfo{editor}{\bibfnamefont{D.}~\bibnamefont{S\'en\'echal}},
  \bibinfo{editor}{\bibfnamefont{A.}~\bibnamefont{Tremblay}}, \bibnamefont{and}
  \bibinfo{editor}{\bibfnamefont{C.}~\bibnamefont{Bourbonnais}}
  (\bibinfo{publisher}{Springer, New York}, \bibinfo{year}{2004}),
  \bibinfo{note}{\text{Chap}. 6}.

\bibitem[{\citenamefont{Kugler and von Delft}(2018)}]{Kugler2018}
\bibinfo{author}{\bibfnamefont{F.~B.} \bibnamefont{Kugler}} \bibnamefont{and}
  \bibinfo{author}{\bibfnamefont{J.}~\bibnamefont{von Delft}},
  \bibinfo{journal}{Phys. Rev. Lett.} \textbf{\bibinfo{volume}{120}},
  \bibinfo{pages}{057403} (\bibinfo{year}{2018}).

\bibitem[{\citenamefont{Kiese et~al.}(2022)\citenamefont{Kiese, M\"uller,
  Iqbal, Thomale, and Trebst}}]{Kiese2021}
\bibinfo{author}{\bibfnamefont{D.}~\bibnamefont{Kiese}},
  \bibinfo{author}{\bibfnamefont{T.}~\bibnamefont{M\"uller}},
  \bibinfo{author}{\bibfnamefont{Y.}~\bibnamefont{Iqbal}},
  \bibinfo{author}{\bibfnamefont{R.}~\bibnamefont{Thomale}}, \bibnamefont{and}
  \bibinfo{author}{\bibfnamefont{S.}~\bibnamefont{Trebst}},
  \bibinfo{journal}{Phys. Rev. Research} \textbf{\bibinfo{volume}{4}},
  \bibinfo{pages}{023185} (\bibinfo{year}{2022}).

\bibitem[{\citenamefont{Taranto et~al.}(2014)\citenamefont{Taranto,
  Andergassen, Bauer, Held, Katanin, Metzner, Rohringer, and
  Toschi}}]{Taranto2014}
\bibinfo{author}{\bibfnamefont{C.}~\bibnamefont{Taranto}},
  \bibinfo{author}{\bibfnamefont{S.}~\bibnamefont{Andergassen}},
  \bibinfo{author}{\bibfnamefont{J.}~\bibnamefont{Bauer}},
  \bibinfo{author}{\bibfnamefont{K.}~\bibnamefont{Held}},
  \bibinfo{author}{\bibfnamefont{A.}~\bibnamefont{Katanin}},
  \bibinfo{author}{\bibfnamefont{W.}~\bibnamefont{Metzner}},
  \bibinfo{author}{\bibfnamefont{G.}~\bibnamefont{Rohringer}},
  \bibnamefont{and} \bibinfo{author}{\bibfnamefont{A.}~\bibnamefont{Toschi}},
  \bibinfo{journal}{Phys. Rev. Lett.} \textbf{\bibinfo{volume}{112}},
  \bibinfo{pages}{196402} (\bibinfo{year}{2014}).

\bibitem[{\citenamefont{Rohringer et~al.}(2018)\citenamefont{Rohringer,
  Hafermann, Toschi, Katanin, Antipov, Katsnelson, Lichtenstein, Rubtsov, and
  Held}}]{Rohringer2018}
\bibinfo{author}{\bibfnamefont{G.}~\bibnamefont{Rohringer}},
  \bibinfo{author}{\bibfnamefont{H.}~\bibnamefont{Hafermann}},
  \bibinfo{author}{\bibfnamefont{A.}~\bibnamefont{Toschi}},
  \bibinfo{author}{\bibfnamefont{A.~A.} \bibnamefont{Katanin}},
  \bibinfo{author}{\bibfnamefont{A.~E.} \bibnamefont{Antipov}},
  \bibinfo{author}{\bibfnamefont{M.~I.} \bibnamefont{Katsnelson}},
  \bibinfo{author}{\bibfnamefont{A.~I.} \bibnamefont{Lichtenstein}},
  \bibinfo{author}{\bibfnamefont{A.~N.} \bibnamefont{Rubtsov}},
  \bibnamefont{and} \bibinfo{author}{\bibfnamefont{K.}~\bibnamefont{Held}},
  \bibinfo{journal}{Rev. Mod. Phys.} \textbf{\bibinfo{volume}{90}},
  \bibinfo{pages}{025003} (\bibinfo{year}{2018}).

\bibitem[{\citenamefont{Tan et~al.}(2018)\citenamefont{Tan, Sun, Kong, Zhang,
  Yang, and Liu}}]{tan2018survey}
\bibinfo{author}{\bibfnamefont{C.}~\bibnamefont{Tan}},
  \bibinfo{author}{\bibfnamefont{F.}~\bibnamefont{Sun}},
  \bibinfo{author}{\bibfnamefont{T.}~\bibnamefont{Kong}},
  \bibinfo{author}{\bibfnamefont{W.}~\bibnamefont{Zhang}},
  \bibinfo{author}{\bibfnamefont{C.}~\bibnamefont{Yang}}, \bibnamefont{and}
  \bibinfo{author}{\bibfnamefont{C.}~\bibnamefont{Liu}}, in
  \emph{\bibinfo{booktitle}{International conference on artificial neural
  networks}} (\bibinfo{organization}{Springer}, \bibinfo{year}{2018}), pp.
  \bibinfo{pages}{270--279}.

\bibitem[{\citenamefont{Brunton et~al.}(2016)\citenamefont{Brunton, Proctor,
  and Kutz}}]{SINDYPNAS}
\bibinfo{author}{\bibfnamefont{S.~L.} \bibnamefont{Brunton}},
  \bibinfo{author}{\bibfnamefont{J.~L.} \bibnamefont{Proctor}},
  \bibnamefont{and} \bibinfo{author}{\bibfnamefont{J.~N.} \bibnamefont{Kutz}},
  \bibinfo{journal}{Proceedings of the National Academy of Sciences}
  \textbf{\bibinfo{volume}{113}}, \bibinfo{pages}{3932} (\bibinfo{year}{2016}).

\bibitem[{\citenamefont{Paszke et~al.}(2019)\citenamefont{Paszke, Gross, Massa,
  Lerer, Bradbury, Chanan, Killeen, Lin, Gimelshein, Antiga
  et~al.}}]{Paszke2019}
\bibinfo{author}{\bibfnamefont{A.}~\bibnamefont{Paszke}},
  \bibinfo{author}{\bibfnamefont{S.}~\bibnamefont{Gross}},
  \bibinfo{author}{\bibfnamefont{F.}~\bibnamefont{Massa}},
  \bibinfo{author}{\bibfnamefont{A.}~\bibnamefont{Lerer}},
  \bibinfo{author}{\bibfnamefont{J.}~\bibnamefont{Bradbury}},
  \bibinfo{author}{\bibfnamefont{G.}~\bibnamefont{Chanan}},
  \bibinfo{author}{\bibfnamefont{T.}~\bibnamefont{Killeen}},
  \bibinfo{author}{\bibfnamefont{Z.}~\bibnamefont{Lin}},
  \bibinfo{author}{\bibfnamefont{N.}~\bibnamefont{Gimelshein}},
  \bibinfo{author}{\bibfnamefont{L.}~\bibnamefont{Antiga}},
  \bibnamefont{et~al.}, in \emph{\bibinfo{booktitle}{Adv. Neural Inf. Process.
  Syst.}} (\bibinfo{year}{2019}), vol.~\bibinfo{volume}{32}, ISSN
  \bibinfo{issn}{10495258}, \eprint{1912.01703}.

\bibitem[{\citenamefont{Harris et~al.}(2020)\citenamefont{Harris, Millman,
  van~der Walt, Gommers, Virtanen, Cournapeau, Wieser, Taylor, Berg, Smith
  et~al.}}]{NumPy}
\bibinfo{author}{\bibfnamefont{C.~R.} \bibnamefont{Harris}},
  \bibinfo{author}{\bibfnamefont{K.~J.} \bibnamefont{Millman}},
  \bibinfo{author}{\bibfnamefont{S.~J.} \bibnamefont{van~der Walt}},
  \bibinfo{author}{\bibfnamefont{R.}~\bibnamefont{Gommers}},
  \bibinfo{author}{\bibfnamefont{P.}~\bibnamefont{Virtanen}},
  \bibinfo{author}{\bibfnamefont{D.}~\bibnamefont{Cournapeau}},
  \bibinfo{author}{\bibfnamefont{E.}~\bibnamefont{Wieser}},
  \bibinfo{author}{\bibfnamefont{J.}~\bibnamefont{Taylor}},
  \bibinfo{author}{\bibfnamefont{S.}~\bibnamefont{Berg}},
  \bibinfo{author}{\bibfnamefont{N.~J.} \bibnamefont{Smith}},
  \bibnamefont{et~al.}, \bibinfo{journal}{Nature}
  \textbf{\bibinfo{volume}{585}}, \bibinfo{pages}{357} (\bibinfo{year}{2020}).

\bibitem[{\citenamefont{Virtanen et~al.}(2020)\citenamefont{Virtanen, Gommers,
  Oliphant, Haberland, Reddy, Cournapeau, Burovski, Peterson, Weckesser, Bright
  et~al.}}]{SciPy}
\bibinfo{author}{\bibfnamefont{P.}~\bibnamefont{Virtanen}},
  \bibinfo{author}{\bibfnamefont{R.}~\bibnamefont{Gommers}},
  \bibinfo{author}{\bibfnamefont{T.~E.} \bibnamefont{Oliphant}},
  \bibinfo{author}{\bibfnamefont{M.}~\bibnamefont{Haberland}},
  \bibinfo{author}{\bibfnamefont{T.}~\bibnamefont{Reddy}},
  \bibinfo{author}{\bibfnamefont{D.}~\bibnamefont{Cournapeau}},
  \bibinfo{author}{\bibfnamefont{E.}~\bibnamefont{Burovski}},
  \bibinfo{author}{\bibfnamefont{P.}~\bibnamefont{Peterson}},
  \bibinfo{author}{\bibfnamefont{W.}~\bibnamefont{Weckesser}},
  \bibinfo{author}{\bibfnamefont{J.}~\bibnamefont{Bright}},
  \bibnamefont{et~al.}, \bibinfo{journal}{Nat. Methods}
  \textbf{\bibinfo{volume}{17}}, \bibinfo{pages}{261} (\bibinfo{year}{2020}).

\bibitem[{\citenamefont{Hunter}(2007)}]{Matplotlib}
\bibinfo{author}{\bibfnamefont{J.~D.} \bibnamefont{Hunter}},
  \bibinfo{journal}{Comput. Sci. Eng.} \textbf{\bibinfo{volume}{9}},
  \bibinfo{pages}{99} (\bibinfo{year}{2007}).

\bibitem[{\citenamefont{Cournapeau et~al.}(2011)\citenamefont{Cournapeau,
  Pedregosa, Michel, Grisel, Blondel, Prettenhofer, Weiss, Vanderplas,
  Cournapeau, Pedregosa et~al.}}]{sklearn}
\bibinfo{author}{\bibfnamefont{D.}~\bibnamefont{Cournapeau}},
  \bibinfo{author}{\bibfnamefont{F.}~\bibnamefont{Pedregosa}},
  \bibinfo{author}{\bibfnamefont{V.}~\bibnamefont{Michel}},
  \bibinfo{author}{\bibfnamefont{O.}~\bibnamefont{Grisel}},
  \bibinfo{author}{\bibfnamefont{M.}~\bibnamefont{Blondel}},
  \bibinfo{author}{\bibfnamefont{P.}~\bibnamefont{Prettenhofer}},
  \bibinfo{author}{\bibfnamefont{R.}~\bibnamefont{Weiss}},
  \bibinfo{author}{\bibfnamefont{J.}~\bibnamefont{Vanderplas}},
  \bibinfo{author}{\bibfnamefont{D.}~\bibnamefont{Cournapeau}},
  \bibinfo{author}{\bibfnamefont{F.}~\bibnamefont{Pedregosa}},
  \bibnamefont{et~al.}, \bibinfo{journal}{J. Mach. Learn. Res.}
  \textbf{\bibinfo{volume}{12}}, \bibinfo{pages}{2825} (\bibinfo{year}{2011}).

\end{thebibliography}

\end{document}